# The color distribution in the Edgeworth-Kuiper Belt [1]

A. Doressoundiram[1], N. Peixinho[1,2], C. de Bergh[1], S. Fornasier[1,3], Ph. Thébault[1], M.A. Barucci[1] and C. Veillet[4]
[1] LESIA, Observatoire de Paris, F-92195 Meudon Principal Cedex, France
[2] Centro de Astronomia e Astrofisica da Universidade de Lisboa, PT-1349-018 Lisboa, Portugal
[3] Dipartimento di Astronomia, Vic. dell'Osservatorio 5, I-35122 Padova, Italy
[4] Canada-France-Hawaii Telescope Corporation, PO Box 1597, Kamuela Hi-96743, USA

Manuscrit pages: 20

Tables: 7

Figures: 11

---

[1] based on observations carried out at the Canada-France-Hawaii Telescope (CFHT), operated by the National Research Council of Canada, the Centre National de la Recherche Scientifique of France and the University of Hawaii.

**The color distribution in the Edgeworth-Kuiper Belt**

Correspondence and requests for materials should be addressed to
A. Doressoundiram
LESIA, Observatoire de Paris, 92195 Meudon Principal Cedex, France
e.-mail: Alain.Doressoundiram@obspm.fr

Abstract

**We have started since 1997 the Meudon Multicolor Survey of Outer Solar System Objects with the aim of collecting a large and homogeneous set of color data for Trans-Neptunian and Centaurs objects. We present here our latest B-V, V-R and R-I colors measurements obtained with the CFH12K mosaic camera of the 3.6m Canada-France-Hawaii Telescope (CFHT). With the colors of 30 objects reported in this work, we have a combined sample of 52 B-R color measurements for 8 Centaurs, 22 Classicals, 13 Plutinos, 8 Scattered objects and 1 object with unidentified dynamical class. This dataset is the largest single and homogeneous published dataset to date, and is large enough to search for compositional structures, interrelations between dynamical classes of objects and correlations with physical and orbital parameters.**

**The color-color diagrams show that all the classes of objects share the same wide color diversity. No significant correlations are seen for the whole population of TNOs and Centaurs, as well as for individual sub-populations, except for the Classicals. Indeed, we found a significant and strong correlation of the colors of Classicals with inclination, eccentricity and perihelion, but nothing with semi-major axis and absolute magnitude. Most of these results are common to other previous works and do not seem to be due to sampling bias. Moreover, a strong correlation with mean excitation velocity [$(V_k(e^2+i^2)^{1/2}$] points toward a space weathering/impact origin for the color diversity. However, thorough modeling of the collisional/dynamical environment in the Edgeworth-Kuiper belt needs to be done in order to confirm this scenario. We found also that the Classical TNOs consist in the superposition of two distinct populations: the dynamically Cold Classical TNOs (red colors, low *i*, small sizes) and the dynamically Hot Classical TNOs (diverse colors, moderate and high *i*, larger sizes). Furthermore, the latter population displays a strong correlation between color and mean excitation velocity. The dynamically Cold Classical TNOs may be primordial while the dynamically Hot Classical TNOs, whose surfaces' colors may be the result of space weathering/impact processes, have possibly been injected from the inner regions of the disk.**

**Our specific observation strategy to repeat color measurements with no rotation artifacts have permitted us to highlight a few objects suspected to have true compositional and/or texture variation on their surfaces. These TNOs are 1998 $HK_{151}$, 1999 $DF_9$, 1999 $OY_3$, 2000 $GP_{183}$, 2000 $OK_{67}$, and 2001 $KA_{77}$ and should be prime targets for further observations in order to study and confirm the color variation with the rotation. Finally, our survey has also highlighted some peculiar objects such as 1998 $SN_{165}$ whose colors and dynamical properties put this object in a new dynamical class distinct from the Classicals, its previously assigned dynamical class.**

# 1 Introduction

The studies of the newly discovered population of the trans-Neptunian objects (TNOs) carried out since 1992 (date of the discovery of the first object; see Jewitt and Luu, 1993) have led to several surprises. First, the dynamical structure of the Edgeworth-Kuiper Belt appears to be much more complex than expected. One can distinguish three main categories of objects: the resonant objects, the scattered disk objects and the classical objects. Most of the resonant objects are in the 2:3 resonance with Neptune, as is Pluto and have been called Plutinos. The scattered disk objects have highly eccentric and inclined orbits. Most of them have a perihelion close to Neptune's orbit. Classical objects have semi-major axes mainly between 40 and 48 AU and, although these objects are far from Neptune's gravitational influence, many of them have relatively high orbital eccentricities and inclinations. From the set of orbits determined so far, it is clear that objects in the Belt have been submitted to a complex dynamical evolution.
According to Duncan et al. (1995) the gravitational influence of Neptune and, to a lesser extent of Uranus and the other giant planets, cannot explain the observed dynamical structure. Other effects have been considered such as planetary migration, temporary presence of large remnants bodies of giant planets formation that went through the Belt before being ejected out of the Solar System, a star that came very close at some point of the evolution of the solar system,….None of the explanations considered so far can explain by itself satisfactorily the observations. Attempts to combine some of these effects or considering others are underway (see e.g. Gladman et al. 2001 for a review).

In addition, it is now generally believed that what we presently see in the Edgeworth-Kuiper Belt is only a very small fraction of the material originally present. A large part of the mass has been lost (see Stern and Colwell, 1997). With a much more massive belt in the past, collisions must have played a very important role, creating small bodies, bringing some "fresh" (protected from radiation) material from inside, modifying orbits, and eventually facilitating the expulsion of bodies outside the belt.

Another main surprise is the large diversity in the colors of the objects (see Barucci et al. 2001). Colors vary from slightly blue to very red. Since it is believed that the TNOs were formed more or less at the same time and in the same cold region of the outer solar system, this important color variation may be mainly due to different degrees of surface alteration.

Looking at the surface of objects belonging to the different populations can help us in retracing their dynamical evolution, environmental and collisional history, and in identifying eventually objects that have not suffered any collisional processing (surfaces affected only by aging). From the photometric studies carried out so far, some trends have been noticed in initial works (Tegler and Romanishin, 2000; Doressoundiram et al. 2001, hereafter DOR01), and subsequent papers (Trujillo and Brown, 2002, Hainaut and Delsanti, 2002). There are mainly an excess of red objects for perihelion distances greater than 40 AU, an indication that highly inclined classical objects are preferentially gray, a wider range of colors for Plutinos than for the other classes of objects. However, there is a clear need for more data in each individual class of objects to confirm these trends and eventually look for other correlations.

We have engaged in an important observational program aimed at collecting a set of high-accuracy photometric data as homogeneous and as large as possible that includes not only objects in the different dynamical classes of trans-Neptunians, but also Centaurs. Centaurs probably constitute a population intermediate between the trans-Neptunians and the Jupiter-family comets. In this paper we discuss our observations of 30 objects that have been carried out at the 3.6 m CFH Telescope in 2000 and 2001. We have combined this new data set with results from previous observations (29 objects) by our group (Barucci et al., 1999, 2000: Doressoundiram et al., 2001) made with similar observational and reduction techniques. We thus have at our disposal a combined data set of 52 B-R colors (and not 59 because of duplicated measurements of the same object or in few cases missing B-V measurements) to look for correlations with various parameters.

## 2 Observations and data reduction

### *2.1 Observations*

This photometric program was part of a granted large program on the Canada-France-Hawaii Telescope dedicated to the discovery/recovery/photometry of TNOs. This multifaceted program was aimed at a dynamical and compositional characterization of the Edgeworth-Kuiper belt, while optimizing the use of the large field of view of the CFH12K camera.

Visible observations were performed during three runs: 2000 December 21-24, 2001 June 26-28 and 2001 August 12-14. We used the 3.6m Canada-France-Hawaii Telescope (CFHT, Mauna Kea, Hawaii) equipped with the CFH12K panoramic CCD camera which is a mosaic of twelve 2Kx4K CCD devices, covering a field of view of 42 by 28 square arc minutes with 0.2 arc second per pixel. Frames were taken through Mould BVRI broadband filters. Two nights (2001 June 28 and 2001 August 13) were not clear, so no photometry was done. All the other nights were photometric, under good seeing conditions.

Objects were selected from their observability, their brightness ($M_V < 22.5$ as estimated by the Minor Planet Center ephemeris service[2]), their position accuracy (e.g. multi-opposition objects), the need to have a new or improved color measurement, and the absence of bright field stars in the vicinity of the object that would compromise the photometric measurement. In addition, newly discovered objects with large orbital uncertainties that were first observed in the recovery part of the large program have been targeted for photometric measurements. The selected objects and their observational circumstances are reported in Table 1. The telescope and instrument characteristics are described in Table 2.

We adopted a specific observation strategy adapted to the variable and faint nature of the trans-Neptunian and Centaur objects. All target objects and standard stars were placed within the CCD #3 (HiRho type CCD of the mosaic) because it was found to be the best in terms of quantum efficiency, charge transfer efficiency and cosmetics. Furthermore, this CCD part exhibits a very low fringing of 0.5% in the I band due to its higher thickness (Cuillandre et al. 2000). The exposure times did not exceed 600 seconds in order to minimize trailing of the TNOs relative to the stars. At opposition, TNOs' motions at 50 AU are roughly 2.6"/hr, thus producing a trail of ~0.4" in 600 seconds, which is small compared to the 0.8"-1.2" FWHM seeing for most of our observations. The telescope was tracked at sidereal rate. Since we aimed at a Signal-to-Noise Ratio (SNR) between 20-30 in all filters, for the faintest objects we co-added images to achieve sufficient SNR. To eliminate systematic errors in the colors (i.e. B-V, V-R, V-I) caused by rotational lightcurve variation, we adopted the following photometric sequence: R-V-B-I-V. This sequence was repeated 2 times during the night in order to i) secure the measurement and then reduce the uncertainty in the color indices; ii) eventually monitor any color variation on the TNO's surface. We think that the issue on the color variability is an important one which is not well addressed in the literature. Indeed, the resurfacing hypothesis (Jewitt & Luu, 2001) predicts such azimuthal color variation on individual objects. Finally, we gave a great attention to the photometric calibration of our data in order to maintain a low level of uncertainty in the final color indices. This goal was achieved by observing a large number of Landolt standard fields (Landolt 1992) over a wide range of airmasses and colors. Transformation equations were solved for zero point, extinction and color terms, leading to a total calibration error typically around 0.01-0.02 mag.

[2] http://cfa-www.harvard.edu/iau/MPEph/MPEph.html

### *2.2 Data reduction*

The images were processed using both MIDAS and IRAF data reduction packages. Data were reduced in the usual manner. First, the frames were bias-subtracted and flat-fielded by a median of the twilight flats. Each image was visually inspected for cosmic rays or bad pixels, and the photometry of the objects was performed using in parallel both aperture correction and large aperture (aperture size of 6xFWHM). All the details of the photometric reduction steps are described in previous papers (Barucci et al., 2000; Doressoundiram et al., 2001) and the interested reader may find all the details about the aperture correction method used. The basis of this method is that we made the photometric measurements by using a small aperture of the order of the size of the seeing disk. Then, to correct for the missing light, we used a mean PSF (i.e. Point Spread Function) built from an average of the field stars of the frame. Several small apertures were used to check for consistency. For all the photometry, the sky value was computed as the mode of sky areas surrounding the object. The advantages in the use of a small aperture are i) decrease the contribution of the sky which could be important and critical for faint objects, ii) minimize the probability of contamination from unseen background sources. We see two main limitations to this method: firstly, each imaging instrument has image distortion across the field which is modulated by seeing and is color dependent. We overcame this problem by analyzing in the image as many bright isolated stars as possible and preferentially stars in the vicinity of the object. Stars with too deviating PSFs (i.e. resolved galaxies or saturated/contaminated stars) were automatically rejected. This results in a final mean PSF calculated with several tens of stars that serves for the correction of small aperture measurements. The distortion effect is thus reflected in the standard deviation of the fit which in turn contribute to the photometric error. Secondly, the TNO and stars do not have exactly the same PSF since usually our images are slightly trailed for the TNOs. However, depending on the seeing and trailing conditions, one can choose an aperture large enough to minimize this error (less than 0.01 mag). Of course, too trailed objects will make the aperture correction technique not applicable. In conclusion, performing the photometry with the aperture correction technique offers large improvements in the accuracy and in the reliability of the measurement, provided that one checks and takes into account the limitations of the method.

## 3 Results

For each individual B, V, R, I magnitude obtained, one sigma uncertainties are based on the combination of several uncertainties. The photometric uncertainty ($\sigma_{pho}$) is based on photon statistics and sky noise. The uncertainty on the aperture correction ($\sigma_{ap}$) is determined from the dispersion among measurements of the different field stars. The final uncertainty in the magnitude is derived from:

$$\sigma = (\sigma_{pho}^2 + \sigma_{ap}^2 + \sigma_{cal}^2)^{1/2}$$

where $\sigma_{cal}$ is the calibration error.

A mean V magnitude is listed on Table 3 and 4, column 2. It is an average of several single V (4 in general) measurements made through the photometric sequences R-V-B-I-V. Geometrical effects were removed by reducing the photometry to H magnitude following Bowell et al. (1989). A canonical value of G = 0.15 was assumed throughout. G is the slope parameter, indicative of the gradient of the phase curve. We have also computed from the derived H magnitude the estimated diameter of the objects (last column of Table 4), assuming an albedo of 0.04 common for dark objects and cometary nuclei. One should be aware of the fact that the sizes are purely indicative and are largely uncertain. For instance, if we used, instead, an albedo of 0.14 (i.e. the albedo of the Centaur 2060 Chiron), all these size estimates would have to be divided by about two. Color indices B-V, V-R, V-I were computed using the closest in time V magnitude measurement in the photometric sequence. When repeated color indices were obtained for the same object, they were found, for the vast majority, to be consistent within the error bars. However, there were some exceptions that will be discussed below. Thus, a weighted mean has been computed, which is listed in Table 3, for each color index (shaded line). Table 4 summarizes for each object the combined color indices, and the absolute magnitude ($H_V$) and size.

By performing two (in most of our observations) photometric sequences over a significant time span, thus obtaining color indices corresponding to different parts of the surface of the rotating object, we could monitor any possible color variation. This occurred a few times, where the differences between the measurements were significant in comparison with the uncertainties. We checked that these differences were not due to reduction or observational (contamination) problems. Indeed, in some cases, the differences were larger (0.1-0.2 mag.) than expected from unseen background sources. However, although improbable, we could not exclude fast rotation of elongated TNOs that would introduce rotational effects within a single photometric sequence. The objects that show color variations are: 1998 $HK_{151}$, 1999 $DF_9$, 1999 $OY_3$, 2000 $GP_{183}$, 2000 $OK_{67}$, and 2001 $KA_{77}$. The color variation could originate from true compositional and/or texture variation on the surface of the object. Here are the details for each object for which color variations have been found.

1998 $HK_{151}$: This Plutino is pointed out as a possible variable object because we obtain a B-V color different from a previous measurement (DOR01). In DOR01, the authors noticed the unusual bluish B-V color of this object (B-V = 0.51 ± 0.09, V-R = 0.43 ± 0.08, R-I = 0.36 ± 0.07) and pointed the need to confirm this measurement. One year later, Boehnhardt et al (2001) confirm the V-R and R-I color measurements but unfortunately did not make a B-V color measurement. In this paper, we report new BVRI colors. The V-R and R-I colors are in good agreement with our previous measurements while the B-V is much higher (B-V = 0.72 ± 0.05). We believe that this last measurement corresponds to the true physical nature of 1998 $HK_{151}$.

1999 $DF_9$: We report first color measurements for this object. This Classical object had a variable V-I color (~3σ difference) over a 40 minutes time span while the other colors are constant. The timeline is relatively short. This V-I color needs to be confirmed before any tentative interpretation is done.

1999 $OY_3$: This Classical object had both V-R and V-I colors variable (>3σ) over a nearly one hour time span. On the other hand, the two B-V colors are the same. The colors are varying between slightly blue colors to almost solar colors. This variation at longer wavelength appears real. The B-V and V-R colors reported by Tegler and Romanishin (2000) are similar to the second set of measurements. Also Hainaut and Delsanti (2002) suggest that this object may host intrinsic activity.

2000 $GP_{183}$: This Plutino had only a variable V-I color (>2σ) over a nearly one hour time span. This is the first color measurement published for this object. This suspicious marginal variation needs further confirmation

2000 $OK_{67}$: This Classical object had a variable B-V color (~2σ) over 45 minutes. Another color measurement for this object has been reported by Delsanti et al (2001). Their B-V color are very close to one of our two measurements. Their V-R color is in good agreement within the error bars with our measurements. The B-V color needs to be measured again in order to confirm any true variability.

2001 $KA_{77}$: We report first color measurements for this object. This Classical object had all B-V, V-R, V-I colors variable (~2-3σ) over a relatively short time span (~23 minutes). If true, this variation may indicate a relatively fast rotation of an elongated object.

# 4 Trends and color properties of Centaurs, Plutinos, Classicals and scattered objects

## *4.1 The color diversity*

We plot (Figure 1) our new data in the now classical B-V versus V-R color diagram, along with all other data published by our group (Barucci et al., 1999, 2000; DOR01, see Table 5). A few objects (e.g. 1998 $VG_{44}$) were observed again during this last campaign. The colors obtained were found very similar to our previous measurements (except 1998 $HK_{151}$ whose case has been discussed in previous section). Thus we now have at our disposal a grand total of 52 objects, the largest published dataset obtained so far by a single team. It is important to note that our analysis is based on an homogeneous dataset (same team, same observation strategy and data reduction techniques), which has not been the case for recent TNOs colors' analyses which used compilations of different datasets. Obviously, with an homogeneous dataset, one can prevent possible inconsistencies between color measurements of the same object from multiple investigators. The reasons for these possible inconsistencies are mainly different observation strategies, filter transformations and/or data processing methods.

With such a significant contribution, the picture of the wide color diversity of TNOs and Centaurs is further outlined. Figure 2 is the same as Figure 1, but shows instead the different populations: 8 Centaurs, 22 Classicals, 13 Plutinos, 8 Scattered objects and 1 object with unidentified dynamical class (see discussion on 1998 $SN_{165}$ below). No distinct behavior is apparent between Centaurs, Classical, Plutino or Scattered objects. Instead, objects in each population are showing a wide spread of colors from gray (solar colors) to very red. This wide range of colors is thought to be the result of concomitant processes acting on the surfaces: the reddening of surface material by irradiation and resurfacing effects by cratering impacts and/or intrinsic activity. However, one has to keep in mind that the weathering/impact resurfacing hypothesis is merely a suggestion, and that no quantitative and precise modeling has demonstrated yet that this suggestion might be true.

The color diversity of TNOs and Centaurs is also present in the two other color-color diagrams (Figures 3 and 4). Colors are mutually correlated (for example, $r_{corr}$=0.79 between B-V and V-R or $r_{corr}$=0.47 between B-V and R-I) showing that a same coloring process is responsible for the reddening from the B (0.43μm) to the I (0.82μm) wavelengths. The lower correlation between B-V and R-I results from the fact that the spectrum of the reddest objects generally flattens toward the infrared. This seems in agreement with the hypothesis that the surfaces of TNOs possess irradiated icy crust. Indeed, as already noted by Hainaut and Delsanti (2002), following the laboratory work of Thompson et al. (1987) on irradiated frosts, the reddest TNOs are expected to have a spectrum that flattens towards the infrared.

This wide color diversity is peculiar to the outer solar system bodies and is not observed among asteroids, comets' nuclei, or planets' satellites. This color diversity is an observational fact that is widely accepted by the community (e.g. DOR01; Jewitt et al. 2001; Delsanti et al. 2001 and references therein). Colors are ranging continuously from gray to very red. However, Tegler and Romanishin (1998, 2000) found instead that the color distribution is bimodal. DOR01 has shown that differences in color measurements among our data and those of Tegler and Romanishin are not the causes of this interpretation: we *do* have color agreement. The origin in this apparent paradox seems to reside in small number statistics and small error bars from Tegler and Romanishin's colors. We strongly encourage observers to better reduce their uncertainties and target those specific objects that lie between the two hypothetical groups.

### *4.2 Correlations*

Figure 5 and 6 are plots showing altogether color, size and orbital elements of outer solar system objects from our survey. We used the same type of representation as first presented in DOR01. Figure 5 and 6 show B-R colors of TNOs and Centaurs in orbital eccentricity (*e*) versus semi-major axis (*a*) and orbital inclination (*i*) versus semis-major axis (*a*), respectively. The B-R color index measures the ratio of the surface reflectance at B (~430 nm) and R (~660 nm) wavelengths. A color palette has been adopted to scale the color spread for objects of our survey from B-R=1.01 (coded as dark blue) to B-R=1.88 (coded as red). In comparison, B-R=1.03 for the Sun, 1.2-1.3 on average for a typical short period comet and 1.97 for the Centaur 5145 Pholus (the reddest known object in the solar system). The size of the symbols are proportional to the corresponding object's diameter (assuming a constant albedo of 0.04.). The advantage of this representation is that it offers to the eye the global color distribution of the Edgeworth-Kuiper Belt and may shed light on some singularities and trends in the belt.
We took the TNOs and Centaurs orbital elements from the Minor Planet Center (Maarsden, 2002). Before we analyze our results, we wish to caution the reader on the relative uncertainty of TNOs'orbital parameters whose accuracy is routinely improved as more data are collected. As a result of the very long orbital period of TNOs (longer than 250 years), the orbital elements cannot be reliably determined in less than three oppositions (Petit et al., 2001). For instance, orbital elements may be dramatically wrong if they are determined from short arcs (months). In the framework of this work, most of our objects have been observed for at least three oppositions, following our observational strategy (see section 2.1), to securely place the objects within the chip #3. So we expect that the figures shown will not change drastically (only slight differences in *a* and *e* for the most recently discovered objects could occur).
On the other hand, if any trends are apparent in color-orbital parameters distributions, and that outliers are obvious, they may be diagnostic of preliminary uncertain orbital elements calculations (see below).
In this paragraph, in a first step, we are analyzing visually Figures 5 and 6, before making statistical tests in the next section. Here are some interesting patterns which come out from these color maps, most of them having already been reported in DOR01, but which are further pronounced with this larger sample of 52 B-R color sample:

1. The eight Centaurs in our sample seem to have redder colors at higher eccentricity. In fact, this trend seems to be common to all objects whose semi major axis is below the 2:3 resonances ($a$<39 UA). This includes the unclassified object 1998 $SN_{165}$ ($a$=37.9, $e$=0.05)
2. Objects with perihelion distances around and beyond 40 AU are *mostly* very red. This characteristic was originally pointed out by Tegler and Romanishin (2000). Classical objects (mostly between the 2:3 and 1:2 resonances) with high eccentricity and inclination are preferentially gray/slightly red, suggesting that some activity (e.g. collisions?) has efficiently rejuvenated (e.g. bluishing) the surfaces in that region of the Edgeworth-Kuiper Belt. Moreover, there is apparently a red color-low inclination cluster of TNOs
3. Contrary to Classical objects, no clear trend is obvious for Scattered TNOs ($a$ > 50 UA). Actually scattered TNOs have bluer colors than Classicals, and lack very red objects.
4. Compared to Classicals, Plutinos lack of any trends in their surface colors, suggesting that the process acting in the "main Edgeworth-Kuiper belt" and responsible for the gray color at high inclinations and eccentricities, seems absent or inefficient in resonance locations.

In that big picture where we have depicted global trends, a few objects appear as outliers in Figures 5 and 6:
1998 $SN_{165}$ ($a$=37.9, $e$=0.05, $i$=4.6) is a gray object. The colors are in agreement with other published values. 1998 $SN_{165}$ is currently classified as a Classical object while we consider it rather as an object with unidentified dynamical class because it is located ahead of the 3:2 resonance. As a Classical object 1998 $SN_{165}$ would look peculiar in Figures 5 and 6 because it is "blue" (e.g. gray surface) while located at relatively low $i$ and $e$. Actually, its colors are more similar to the low excited objects located at $a$ < 39 AU (ahead of the 2:3 resonance). Possibly, this object is a member of a rather separate dynamical class, as suggested by Gladman (2002) on dynamical considerations, distinct from the Classical objects. Actually, several objects have been discovered in that part of the KB ($a$~38 UA, low $i$) and are thought to be very primitive objects and stable in their original location since the early stages of the Solar System formation.

Among scattered TNOs, also two objects 1999 $DE_9$ (*a*=56.0, *e*=0.42, *i*=7.6) and 1999 $CC_{158}$ (*a*=54.4, *e*=0.28, *i*=18.7) may be considered marginal as their colors are moderately red compared to the other 6 Scattered TNOs which are gray. Their orbits are no more excited or less excited than the other Scattered TNOs, as to explain their color differences. No color variation has been detected on less than one hour, and our colors are in good agreement with other published values. They are multi-opposition objects, so that their orbital parameters can be considered as secure. Are these two objects real members of the SDO population? Are the Scattered TNOs colors gray in general? Obviously, more data need to be gathered to increase the statistics on the SDO population.
However, these tentative analyses on the outliers should be taken with care, as always when we are dealing with small numbers.

All these interesting traits do indeed show up clearly in Figure 5 and 6. But, in order to assess their significance, we need to perform statistical tests. Contrary to our initial TNOs surveys, we can now perform on this large and homogeneous dataset robust and simple tests in order to determine how reliable any of these results are. Very recently, some reliable correlations have been found, like the inclination-color correlation reported by Trujillo and Brown (2002).
To investigate all the possible relationships between color and physical/orbital characteristics (size, absolute magnitude, orbital elements, …), we used the Spearman rank correlation statistics $r_{corr}$ ( $-1 \leq r_{corr} < 1$) for our dataset. The method (Press et al. 1992) is a nonparametric test. We use a nonparametric test (as the Kolmogorov-Smirnov test that we will use in section 6) because such techniques are not dependent upon an underlying assumption of normal distributions for the sampled variables. Heuristically, instead of using the real data values for its computation, the Spearman method ranks the (x,y) data points as a function of x values and measures how *unranked* the y will be, or vice-versa. One of the advantages of this test is that it makes no assumption about any fitting function to estimate the correlation (as is the case of the Pearson or linear correlation method). The closer to 1 or –1 is $r_{corr}$ the stronger is the correlation between the two variables, while a value close to 0 indicates that they are uncorrelated.
Table 6 summarizes the correlations obtained for a selection of interesting cases. $P(r > r_{corr})$ gives the probability that a correlation coefficient equal to or larger than the one measured could be obtained by chance in an uncorrelated sample. The probability P follows the t-Student distribution independently of the original distribution of our data sample. 1-P gives the confidence level of the correlation coefficient found. For instance, $P(r > r_{corr}) = 0.003$ indicates a confidence level of 99.7% which is the nominal $3\sigma$ criterion for a statistically significant correlation.
First of all, we did not find any kind of correlation when we considered the whole TNO and Centaur populations in our dataset (N=52), as expected from the examination of the color maps. We then looked specifically at each population.

**Centaur objects (N=8):**
We wanted to investigate the trend seen in color maps (point 1 of the above list). That is: the eight Centaurs in our sample seem to have redder colors at higher eccentricity. We found indeed a strong correlation $r_{corr}$=0.62 between B-R and *e*. However, the significance found is low ($1.6\sigma$). Also, we found moderate correlation with *a*, *i*, and aphelion (*Q*), but still not statistically significant. Before speculating on the possible origin of this trend, we need to wait for further data. Actually, when increasing our sample with 8 additional color data found in the literature, all these correlations vanish.

**Plutinos (N=13):**
We found only weak correlations with *e*. Finally, with the extended dataset (see below) of Plutinos' colors, we do not see any correlation at all (see Table 6).

**Scattered TNOs (N=8):**
We confirmed the point 3 of the list. The colors of Scattered TNOs are not correlated with any of *a*, *e*, *i*, size or H. As for the average color, the colors of Scattered TNOs are indeed bluer than those of the other populations. However we have at our disposal only 8 colors in our Scattered TNOs sample. So, in principle, small number statistics could be responsible for this result. Hainaut and Delsanti (2002) results on a larger database of 95 objects, built as the combination of several published datasets, did not find any systematic differences of colors between the different TNO populations.

**Classical objects (N=22):**
For the rest of the paper, we are considering as Classical TNOs the objects having their semi-major axis between 40.5 UA (thus excluding the Plutinos) and roughly 48 AU. By taking these boundaries, we are automatically rejecting 1998 $SN_{165}$ ($a$=37.9 AU) that we have discussed before as a peculiar object which may rather belong to a dynamical class distinct from the Classical TNOs. Actually several TNOs whose orbital parameters are close to 1998 $SN_{165}$ have been discovered in that small region of the EKB ($a$~38 AU, low $i$) predicted to be stable by Duncan et al (1995) and obviously there is a nomenclature problem for the Classical objects posed by this population (see Gladman 2002).
For the Classical TNOs, the most obvious relationship shown in color maps involves inclination. Indeed, we found a strong correlation between color and inclination (Figure 7), but only for the sub-populations of Classical and SDO objects (N=30). Excluding the Scattered objects from the sample gives a better correlation, but is still statistically tentative given the small number of scattered TNOs of our dataset (N=8). For B-V vs. inclination, the Spearman's rank correlation coefficient is found to be $r_{corr}$= -0.80 (N=22); the probability of this $r_{corr}$ or a more significant one occurring in an uncorrelated sample is P=9 $10^{-6}$ (4.4$\sigma$ significance for gaussian statistics). For B-R vs. inclination, we found $r_{corr}$ = -0.72 with a probability of P = 0.0002 (3.8$\sigma$ significance). We note that the correlation with B-V is stronger than with B-R, and also than with B-I ($r_{corr}$=-0.69). This is a result which is general to all the correlations we have found: the correlation is stronger at shorter color wavelengths. This trend is a consequence of what was already noticed in color-color plots (Figures 2-4): most of the reddest TNOs have a spectrum that flattens toward the infrared.
The color-inclination correlation has been reported recently by Trujillo and Brown (2002) on their dataset of N=24 B-R colors including however both Classical and Scattered objects. Their trend corresponded to 3.1$\sigma$ significance level. On the other hand, Jewitt and Luu (2001) did not find any correlation with color in their sample of 28 B-I color indices. We attribute this to the high proportion of resonant objects included in their sample which hid the correlation.
How reliable is the color-inclination correlation regarding the orbital uncertainties? Actually, the inclination is generally the best determined of the six orbital elements because it is uniquely calculated by the motion of the TNO perpendicular to the ecliptic. Even a short arc of observation at opposition (where most TNOs are discovered) is sufficient to obtain *i* with an accuracy of less than 0.5°.
We also investigated the correlation with eccentricity. We also found a strong one. We got a B-R vs. eccentricity correlation coefficient $r_{corr}$= -0.60; with a probability of P=0.003 (3$\sigma$).

Thus, the correlation found is statistically reliable. Such a correlation has been only reported by Hainaut and Delsanti (2002).
Looking for other correlations with color, we found only one with perihelion distance (Figure 8) We got a B-R vs. perihelion correlation statistic $r_{corr}$=0.76; with a 4.1σ significance for the Classical objects (N=22). Trujillo and Brown (2002) also noticed this correlation, but further attributed it to a sampling bias. They based this conclusion on the analysis of a constant *i* subsample in which they found no correlation between color and perihelion (Figure 3 of their paper). We did the *same* analysis by considering a constant *i* subsample consisting of objects between 7° < *i* < 15° (same boundaries as Trujillo and Brown). We obtained a result opposite to that of Trujillo and Brown: indeed we still found a strong correlation $r_{corr}$ =0.74 (N=8, 2.1σ significance) between color and perihelion. However, more data would be preferable, especially at low q to increase the significance, but this test indicates that this trend may be real. (see Figure 9). We suspect that the reason why the above authors did not find a trend is that they included in their subsample 1997 $SZ_{10}$ and 1998 $SM_{165}$ (q~30 AU) which are likely to be located in the 1:2 resonance. These resonant objects, like the Plutinos, might not show any correlation.

We did not find any correlation between color and absolute magnitude (Figure 10) or between color and semimajor axis. These results yield for all sub-populations and for all color indices. The lack of correlation between color and absolute magnitude translates into a lack of correlation between color and size, even if possible differences in albedos between the different objects are taken into account. Indeed we checked that, even if we varied the albedo of the objects between 0.04 and 0.14 (value for Chiron), with the reddest objects having the smallest albedos, as suggested by Fernandez et al., 2002, the main characteristics observed in Figure 10 would remain (large spread of colors for all sizes in each population). Noteworthy, Hainaut and Delsanti (2002), based on the analysis on their combined dataset (N=95) found a trend for Classicals with faint H to be redder than the others. Moreover, they found the opposite trend for the Plutinos (faint H tend to be bluer). We did not find any of these trends in our homogeneous but smaller dataset. Of course, these opposite trends need to be confirmed by a larger observational dataset, and still their interpretation remain difficult.

In order to check that our single dataset was not responsible for the results found and at the same time increase the sample size, we repeated this analysis combining our data with all other major BVR dataset previously published (Tegler and Romanishin, 1998, 2000; Jewitt and Luu, 2001; Delsanti et al. 2001; Trujillo and Brown, 2001; Boehnhardt et al. 2001). For multiple measurements of the same object, we took the mean. However, we should stress that combination of many different datasets may introduce artifacts (see section 4.1). Combining these additional data with our dataset, and selecting only the Classical objects, the significance of the inclination and color correlation increases to 4.0σ (N=50). On the other hand, the correlation of color with eccentricity and perihelion is weak and no more significant.
Nevertheless, we believe that the strong and significant correlations we have found for the Classicals of our survey are real because they were built up from an homogeneous and large enough dataset. And even if some of our measurements may agree within the uncertainties with those of a given observer, we believe that a single and homogeneous dataset is better for a statistical analysis. Of course, the correlations with *e* and particularly with *q* need to be strengthened with additional data. The correlation between color and perihelion is an interesting and important result since it may give some clues on the structure of the EKB (structure in *a* and *e*, extension of the disk…). Obviously, we will continue to increase our observational dataset with additional colors, especially for those TNOs whose orbits lye at small $q$ ($q$~30 AU).

## 5 The TNOs' mean excitation velocity or could the color diversity be the result of collisional processes?

The correlation with $i$ and $e$ for Classical TNOs points out that objects dynamically excited have preferentially less red surfaces. This is an observational evidence that some process is more efficient in that part of the belt than in other parts in reworking the surfaces of TNOs.
We present one more color-correlation estimate, i.e. with the rms excitation:

$$<V_{rms}> = V_k(e^2+i^2)^{1/2}$$

where $V_k$ is the Keplerian orbital velocity given by $V_k$=(29.8 km/s)$a^{-1/2}$, and a is the semi-major axis expressed in Astronomical Units (AU).
$V_{rms}$ is the mean excitation velocity. This parameter is of great interest, since it gives a first order approximation of the *collisional encounter velocity* for a given TNO. Such an information might be useful, because one of the proposed explanations for the color diversity within the belt is the effect of collisional resurfacing after mutual impacts among TNOs (e.g. Luu and Jewitt 1996). This scenario is based on the concomitant action of two time dependent processes: the reddening and darkening of icy surfaces by solar and galactic irradiation, and the excavation of fresh, primordial ices as the results of collisions. These fresh materials, made of brighter and more neutral ices, will thus *bluer* the surface. Of course, the collisional resurfacing hypothesis requires that both processes have about the same timescale. If it is not the case, we will have either a population uniformly red or a population uniformly gray. More recently, Gil-Hutton (2002) obtained similar results with a different resurfacing model. Furthermore, referring to the laboratory work of Thompson et al. (1987), they take into account the possibility that, with further irradiation (after about $6.10^8$ years), the red crust would become again gray in color while retaining its low albedo. This behavior obviously complicates the interpretation of the color distribution without the knowledge of the albedo distribution. Indeed, only the albedo will permit to distinguish between *gray* TNOs whose surface has been extensively reworked by collisions (high albedo) and *gray* TNOs whose surfaces possess a dark thick irradiation mantle (low albedo).

The collisional resurfacing scenario remains hypothetical and is still the subject of great discussions. It presents nevertheless the advantage of making seemingly simple predictions concerning the color correlation within the Edgeworth-Kuiper Belt. Basically, the most excited objects should be the ones most affected by energetic impacts and thus the most gray ones. It is thus very tempting to check the correlation between the color index and $V_k(e^2+i^2)^{1/2}$, since both $i$ and $e$ should contribute to the average encounter velocity of a TNO. The result of this correlation estimation is presented in Table 6. An obvious result is that the correlation is very good (Figure 11) for the Classical objects. The Spearman's rank correlation coefficient is found to be r= -0.77 with a significance of 4.2$\sigma$ (N=22). This result suggests that collisions may play a role in the color diversity. This tends to confirm the results of Stern (2001) obtained with a smaller object sample made of all sub-populations.
One must nevertheless remain very careful when interpreting such correlations. The $V_{rms}$

parameter gives indeed a very partial information. Strictly speaking, it gives only estimates of the TNO proper excitation which might strongly differ from its average impact velocity. This parameter does not take into account the fact that a collision can occur between bodies originating from different parts of the belt and each having a different excitation contributing to the total impact velocity. Furthermore, the $V_{rms}$ parameter is a quantity which gives the average impact velocity in a maxwellian population with average values of $e$ and $i$. It does not give the average impact velocity for a single object with $e$ and $i$. Thus, the predictions of the collisional resurfacing scenario are not as simple as they might seem. An accurate study of the impact velocity distribution must be done using more complex tools. This problem is addressed in Thébault and Doressoundiram (2002) using accurate deterministic numerical simulations which show obvious similarities, but also clear departures from the observed color distribution.

## 6 The red-low inclination cluster or could the color diversity be the result of true compositional variegation?

Levison and Stern (2001) showed that the Classical objects are the superposition of two distinct populations. The first population would contain dynamically hot objects with high inclinations orbits and big objects. The second population would contain dynamically cold objects with low inclination orbits and relatively small TNOs. Independently, Brown (2001), analyzing the unbiased inclination distribution of the Edgeworth-Kuiper belt, similarly concluded to a two-component inclination distribution of the Classical TNOs. Based on this bimodal behavior of the Classical TNOs in both inclination and size, Levison and Stern (2001) speculated that the hot population originated from the inner regions of the disk where the size distribution and color varied with heliocentric distance. On the other hand, the cold population (low $i$, $a$>41 UA) should be primordial, and dynamically stable over the age of the Solar System, according to results of Duncan et al (1995). Members of this cold population should have very similar physical characteristics because they were formed at the same time and within a relatively small region. The existence of two distinct Classical populations appears to be borne out by our results shown in Figure 6 where a red-low inclination cluster of TNOs is apparent.

In order to check that the two populations are statistically different regarding both their colors and sizes, we apply the two-dimensional Kolmogorov-Smirnov (K-S) statistical test (Peacock, 1983). This test computes the probability that two 2D populations are extracted from the same parent population. A zero probability means that the two populations are different while an unit probability means they are the same. To test this hypothesis, we separate the “Hot Classical population” from the “Cold Classical population” of our sample of Classical objects by putting an inclination cutoff at $i = 5°$ (see Table 7), following Levison and Stern assumption. We find that the K-S probability for the two populations characterized both by their colors and size distribution, is 0.01. Thus, the Cold Classical population and the Hot Classical population are most probably different. We check that the K-S result is not affected when varying the sizes, considering that we may have albedo differences. Indeed, the H (absolute magnitude) distribution of the two populations are completely different.

If still speculative, this scenario is however opposite to the collision resurfacing hypothesis suggested by the $V_{rms}$ correlation. In others words, the color diversity could originate from true compositional diversity, and not from collisional processes reworking surfaces of TNOs.

To test this hypothesis, we analyze the color-orbital excitation distribution within the Hot Classical population. Surprisingly, we still find a 3.3σ (N=13) strong and significant correlation between B-R color and mean excitation velocity [$V_k(e^2+i^2)^{½}$)]. This result suggests that the dynamical excitation of TNOs' orbits certainly plays a role in the color diversity seen in the Hot Classical population. Does this result exclude the hypothesis of an origin in the inner disk? We do not think so because the temperature gradient across the Uranus-Neptune region is expected to be small. The temperature gradient between 20 and 30 AU is only 12 K, barely enough to induce strong compositional differences among TNOs. Therefore, the color diversity seen among Hot Classical TNOs could be explained by collisional resurfacing processes as sustained by the color-$V_{rms}$ correlation found.

Given these results and the lines of supporting evidence for a two-component inclination distribution of the Classical TNOs, reported by the above quoted authors, we speculate on a possible structure of the classical population. We confirm the statistical reality of the two populations. The Cold Classical objects consist of small objects at low *i*. Our survey demonstrates that they are also mostly red (except 1999 $HR_{11}$ which is slightly different with a moderately red color. This object was fainter than expected –$M_V$=23.9– and the uncertainties are consequently very large. Its color measurements need to be refined). The Hot Classical objects consist of larger members at higher inclination. We found that their colors are very diverse.

## 7 Conclusions and perspectives.

We reported BVRI colors for 30 trans-Neptunian and Centaur objects obtained at the Canada-France-Hawaii telescope. This observation campaign is part of our ongoing Meudon Multicolor Survey of Outer Solar System Objects. Combining this last dataset with our previous published colors, we obtained a unique and homogeneous dataset of 52 B-R colors, which is the largest dataset obtained so far by a single team. We then analyzed this dataset to look for correlations between colors and various parameters such as heliocentric distance, absolute magnitude or orbital parameters.
The main results of this analysis are as follows:
We confirmed the wide and continuous spread of all B-V, V-R and R-I colors. The different dynamical classes (e.g. Centaurs, Plutinos, Classicals and Scattered) seem to share the same color diversity. We find no correlation between size, colors, or heliocentric distance for the whole TNOs and Centaurs populations of our survey.As a result of our observing procedure of repeating color measurements, we highlighted a few objects for which color variation have been found and thus that may be diagnostic of true compositional and/or texture variation on the surface of the objects. These TNOs are 1998 $HK_{151}$, 1999 $DF_9$, 1999 $OY_3$, 2000 $GP_{183}$, 2000 $OK_{67}$, and 2001 $KA_{77}$ and should be prime targets for further observations in order to study and confirm the color variation with the rotation.

We did not find any kind of correlation for the individual Plutinos, Centaurs or Scattered populations with orbital parameters. The only exception is the Classical objects for which we found a strong and significant correlation of color with orbital eccentricity, orbital inclination and perihelion distance. We also investigated the correlation between color and mean excitation velocity [$V_k(e^2+i^2)^{½}$], and as previously reported in the literature, we found a significant correlation. This is a strong argument for supporting the collisional resurfacing hypothesis (Luu and Jewitt, 1996). However, considering only the mean excitation velocity of a given TNO gives only a partial information on this object's collision dynamics. It does not take into account the fact that this body might collisionaly interact with impactors from very different regions of the disk. Nevertheless, the correlation found is an encouraging result that led Thébault and Doressoundiram (2002) to perform quantitative collisional/dynamical modeling in order to estimate the role of collisions in that region of the EK belt.

Instead of showing a continuous color distribution from low inclined to high inclined orbits, the Classical objects may consist in the superposition of two distinct populations, as suggested by Levison and Stern (2001) and Brown (2001). A supporting line of evidence in our data to this scenario is the presence of a red-low inclined cluster of Classical objects for $i < 5°$. With a two dimensional Kolmogorov-Smirnov test, we found that the dynamically Cold Classical TNOs (red colors, low $i$, small sizes) and the dynamically Hot Classical TNOs (diverse colors, moderate and high $i$, larger sizes) are two populations significantly different that could not be extracted from the same parent population. Furthermore, the Hot Classical TNOs display a strong correlation between color and mean excitation velocity. Based on these results, we suggest that the Cold Classical TNOs is a primordial population and that the Hot Classical TNOs may have been injected from the inner regions of the disk, and that their surfaces' colors may be the result of space weathering/impact processes. However, quantitative collisional/dynamical modeling need to be done, as well as additional observational data need to be obtained in order to resolve this issue.

An analysis based on visible colors with no albedo measurement available (only one TNO, Varuna, has had its albedo measured) has its limits. Important differences in the albedos of TNOs that have the same visible colors may exist. This would be the case, for instance, of objects that have a gray color because of the presence of fresh ices as compared with objects with a gray color because of a very long period of irradiation (see section 5). And putting these two categories of objects in the same class would be erroneous. Furthermore, objects with similar colors in the visible may have very different colors in the near-infrared, which would mean different surface compositions. Unfortunately, because of the faintness of the objects, near-infrared colors have been obtained for only a very limited number of objects. The type of classification presented in this paper has the advantage of being the one for which the largest dataset can be built up. As more albedo measurements become available (in particular with the SIRTF satellite to be launched in 2003) its value will significantly increase.

Finally, from the global trends found with colors and orbital parameters, we highlighted a peculiar object. 1998 $SN_{165}$ ($a$=37.9, $e$=0.05, $i$=4.6) is an object with yet unidentified dynamical class. Based on its gray colors which are atypical for a low excited object, and given the fact that its orbit lies in a very stable region of the belt (Duncan et al. 1995), we concluded that 1998 $SN_{165}$ belongs to a rather new dynamical class, distinct from the Classicals. While the origin of its gray color is unknown and still difficult to interpret without the knowledge of the albedo, it will be very interesting to see if other objects from the same class share the same color properties. Also of high interest, the visible spectrum of this object can be easily obtained with 8-10-meter class telescope. If this object's surface is covered with extremely irradiated ices (e.g. gray dark colors), its spectrum may show a bend in the B region possibly diagnostic of its evolved state (see Hainaut and Delsanti, 2002).

*Acknowledgement*: We thank Solenne Blancho for her help in part of the data reduction. We are very grateful to Brett Gladman for helpful discussions and ideas. Many thanks to R.P. Binzel for a careful reading of the manuscript. We would like also to thank M.E. Brown for fruitful refereeing and comments.

## References

Barucci, M. A., Fulchignoni, M., Birlan, M., Doressoundiram, A., Romon, J., & Boehnhardt, H. 2001. A&A, 371, 1150

Barucci, M.A., Romon, J., Doressoundiram, A., & Tholen, D. 2000, AJ, 120, 496

Barucci, M.A., A. Doressoundiram, M. Fulchignoni, D. Tholen, & M. Lazzarin, 1999. *Icarus*, 142, 476.

Boehnhardt, H., G.P. Tozzi, K. Birkle, O. Hainaut, T. Sekiguchi, M. Vair, J. Watanabe, G. Rupprecht, & The FORS Instrument Team. 2001. *A&A*, 378, 653.

Bowell, E., Hapke, B., Domingue, D., Lumme, K., Peltoniemi, J. & Harris, A.W. 1989. In Asteroids II (R.P. Binzel, T. Gehrels and M.S. Matthews, Eds.), pp. 524-556. Univ. of Arizona Press, Tucson.

Brown, M. E. 2001, AJ., 121, 2804.

Cuillandre, J-C., Luppino, G.A., Starr, B.M., & Isani, S.2000. Proc. SPIE 4008, 1010-1021.

Davies, J.K., S. Green, N.McBride, E. Muzzerall, D.J. Tholen, R.J. Whiteley, M.J. Foster, & J.K. Hillier, 2000. *Icarus*, **146**, 253-262.

Delsanti, A. C., Boehnhardt, H., Barrera, L., Meech, K. J., Sekiguchi, T., Hainaut, O. R. 2001, A&A, 380, 347.

Doressoundiram, A., Barucci, M.A., Romon, J., Veillet, C., 2001, Icarus, 154, 277.

Duncan, M.J., Levison H. F. & Budd, S. M. 1995. AJ., 110, 3073.

Fernández, Y. R., Jewitt, David C. & Sheppard, S. S. 2002. AJ., 123, 1050.

Gil-Hutton, R. 2002, Planet. Space Sci., 57, 62

Gladman, B. 2002. In Highlights of Astronomy, ASP (A. Lemaître & H. Rickman, Eds.), in press.

Gladman, B., Kavelaars, J. J., Petit, J-M., Morbidelli, A., Holman, M. J. & Loredo, T. AJ., 122, 1051.

Hainaut, O. R. & Delsanti, A. C. 2002, A&A, in press.

Jewitt, D., & Luu, J. 2001, AJ, 122, 2099

Jewitt, D. C., Aussel, H., & Evans, A. 2001. Nature, 411, 446

Landolt, A. 1992. AJ., 104, 340-371.

Levison H. F. & Stern, S.A. 2001. AJ., 121, 1730.

Luu, J., & Jewitt, D. 1996, AJ, 112, 2310

Marsden, B. G. 2002. List of Transneptunian and Centaur objects.(April 2001) http://cfa-www.harvard.edu/iau/lists/TNOs.html and http://cfa-www.harvard.edu/iau/lists/Centaurs.html.

Peacock, J.A., 1983, MNRAS, 202, 615-627

Petit, J.-M., Gladman, B., Kavelaars, JJ, Holman, M., Parker, J., Grav, T., & Veillet, Ch. 2001. BASS 33, 1030.

Press, W. H., Teukolsky, S. A., Vetterling, W. T., & Flannery, B. P. 1992, Numerical Recipes in C: The Art of Scientific Computing (2d ed.; Cambridge: Cambridge Univ. Press)

Stern, A., 2001, BASS 33, 1033

Stern, S.A., & J.E. Colwell, 1997. ApJ., **490**, 879-882.

Tegler, S.C. & W. Romanishin, 2000. Nature **407,** 979-981.

Tegler, S.C. & W. Romanishin, 1998. Nature **392,** 49-50.

Thébault Ph. & Doressoundiram, A. 2002, AJ, submitted

Thompson, W. R., Murray, B. G. J. P. T., Khare, B. N., & Sagan, C. 1987. JGR **92,** 14933.

Trujillo, C., & Brown, M. 2002, ApJ, 566, 125.

## Figure captions

**Figure 1**: B-V versus V-R plot. The filled circles represent our latest work while the empty squares are from our previous works (N=52). The star represents colors of the Sun.

**Figure 2**: B-V versus V-R plot. Same sample as figure 1, but showing the different populations. The star represents colors of the Sun

**Figure 3**: B-V versus R-I plot. The whole data set of our TNOs and Centaurs' survey is represented. The star represents colors of the Sun.

**Figure 4**: Same as Fig. 3 in the V-R versus R-I plane.

**Figure 5**: Colors of Centaurs and TNOs in our survey (52 objects) in the orbital eccentricity versus semi major axis plane. The size of the symbols are proportional to the corresponding object's diameter. Colors are scaled from blue (gray objects) to red (very red objects). 2:3 (a~39.5 AU) and 1:2 (a~48 AU) resonances with Neptune are marked as well as the q=40AU perihelion curve.

**Figure 6**: Same as Figure 5 in the orbital inclination versus semi major axis plane.

**Figure 7**: Inclination versus B-R color plot of Classical objects. A correlation exists which yields only for the Classical objects. Spearman's rank correlation statistics gives $r_{corr}$ = 0.72 (3.8$\sigma$ significance). A linear least-squares fit has been plotted to illustrate the correlation.

**Figure 8**: B-R color index versus perihelion distance plot for the objects of our survey, including both Centaurs and TNOs. For comparison, B-R =1.03 for the Sun. No obvious trend is apparent for the whole population (N=52) as well as for individual population except for the Classical TNOs for which a significant 3$\sigma$ correlation between color and perihelion has been found. A linear least-squares fit has been plotted to illustrate the correlation.

**Figure 9**: B-R color index versus perihelion plot for the constant *i* subsample of our dataset. Same figure as Figure 3 (bottom) of Trujillo and Brown (2002). Contrary to the latter authors who argue for a sampling bias, we do find a strong correlation $r_{corr}$ =0.74 (2.1$\sigma$ significance) between color and perihelion in our constant *i* subsample as estimated by the Spearman rank correlation method. A linear least-squares fit has been plotted to illustrate the correlation.

**Figure 10**: B-R color index versus absolute magnitude plot for our survey.

**Figure 11**: B-R color index versus mean excitation velocity plot of Classical objects, showing that $V_k(e^2+i^2)^{½}$ is correlated with color. Spearman's rank correlation statistics gives $r_{corr}$ = -0.77 (4.2$\sigma$ significance). A linear least-squares fit has been plotted to illustrate the correlation.

# Tables

## Table 1: Observational circumstances.

| *Object* | *Group* | *Date* | *Δ (AU)* | *r (AU)* | *α (deg)* |
|---|---|---|---|---|---|
| **10370 Hylonome** | Centaur | 2001 Jun 26 | 18.885 | 19.362 | 2.7 |
| **1995 $SM_{55}$** | Classical | 2000 Dec 21 | 38.904 | 39.389 | 1.3 |
| **1995 $TL_{8}$** | Scattered | 2000 Dec 23 | 41.969 | 42.427 | 1.2 |
| **1996 $TK_{66}$** | Classical | 2001 Aug 14 | 42.288 | 42.938 | 1.0 |
| **1998 $BU_{48}$** | Centaur | 2000 Dec 24 | 26.879 | 27.529 | 1.6 |
| **1998 $HK_{151}$** | Plutino | 2001 Jun 26 | 29.602 | 30.406 | 1.2 |
| **1998 $QM_{107}$** | Centaur | 2001 Aug 12 | 16.914 | 17.919 | 0.4 |
| **1998 $TF_{35}$** | Centaur | 2000 dec 21 | 18.447 | 19.273 | 1.6 |
| **1998 $VG_{44}$** | Plutino | 2000 Dec 23 | 29.468 | 30.303 | 1.0 |
| **1998 $WH_{24}$** | Classical | 2000 Dec 22 | 41.473 | 42.311 | 0.7 |
| **1998 $WW_{24}$** | Plutino | 2000 Dec 24 | 30.851 | 31.613 | 1.1 |
| **1999 $CC_{158}$** | Scattered | 2000 Dec 22 | 41.704 | 42.443 | 0.9 |
| **1999 $CD_{158}$** | Classical | 2000 Dec 22 | 47.465 | 48.312 | 0.6 |
| **1999 $CL_{158}$** | Classical | 2000 Dec 23 | 33.333 | 34.132 | 1.0 |
| **1999 $DE_{9}$** | Scattered | 2000 Dec 22 | 33.526 | 33.923 | 1.5 |
| **1999 $DF_{9}$** | Classical | 2000 Dec 23 | 39.343 | 39.781 | 1.3 |
| **1999 $OX_{3}$** | Centaur | 2001 Jun 27 | 26.077 | 26.935 | 1.2 |
| **1999 $OY_{3}$** | Classical | 2001 Aug 14 | 38.182 | 39.155 | 0.4 |
| **1999 $TD_{10}$** | Scattered | 2000 Dec 22 | 12.314 | 12.444 | 4.5 |
| **1999 $UG_{5}$** | Centaur | 2000 Dec 22 | 7.231 | 8.034 | 4.3 |
| **1999 $XX_{143}$** | Centaur | 2000 Dec 21 | 23.860 | 24.722 | 1.1 |
| **2000 $FE_{8}$** | Scattered | 2000 Dec 24 | 35.543 | 35.448 | 1.6 |
| **2000 $GP_{183}$** | Plutino | 2001 Jun 27 | 36.903 | 37.287 | 1.5 |
| **2000 $OJ_{67}$** | Classical | 2001 Jun 27 | 41.861 | 42.557 | 1.0 |
| **2000 $OK_{67}$** | Classical | 2001 Jun 26 | 40.254 | 40.798 | 1.2 |
| **2000 $QC_{243}$** | Centaur | 2001 Aug 14 | 18.329 | 19.302 | 0.9 |
| **2000 $WR_{106}$ (Varuna)** | Classical | 2000 Dec 21 | 42.078 | 43.045 | 0.2 |
| **2001 $KA_{77}$** | Classical | 2001 Aug 14 | 48.638 | 48.928 | 1.1 |
| **2001 $KD_{77}$** | Plutino | 2001 Aug 12 | 34.826 | 35.163 | 1.6 |
| **2001 $KX_{76}$** | Plutino | 2001 Aug 12 | 42.994 | 43.268 | 1.3 |

Notes: Coordinates are for the date of observations at 00 UT. r, Δ, α are, respectively, the heliocentric distance, the topocentric distance and the phase angle of the object (Minor Planet Ephemeris Service http://cfa-www.harvard.edu/iau/MPEph/MPEph.html)

## Table 2: Telescope/instrument characteristics.

| *Telescope* | *Instruments* | | *Filters* | | |
|---|---|---|---|---|---|
| | *Detector* | *Pixel scale* | *Type* | *$\lambda_c$(nm)* | *fwhm (nm)* |
| **CFHT 3.6m** | CFH12K 12 MIT/LL 2Kx4K | 0.20” | Mould B | 431.2 | 99.0 |
| | | | Mould V | 537.4 | 97.4 |
| | | | Mould R | 658.1 | 125.1 |
| | | | Mould I | 822.3 | 216.4 |

**Table 3: Individual color measurements.**

| ***Object*** | ***V*** | ***Date (UT)*** | ***UT(B) (hr)***[1] | ***B-V*** | ***UT(R) (hr)***[1] | ***V-R*** | ***UT(I) (hr)***[1] | ***V-I*** | ***Var***[2] |
|---|---|---|---|---|---|---|---|---|---|
| **Solar colors** | | | | 0.67 | | 0.36 | | 0.69 | |
| **Hylonome** | 22.59 ± 0.08 | 2001 Jun 26 | 7.267 | 0.81 ± 0.11 | 7.510 | 0.32 ± 0.08 | | | |
| **Hylonome** | | 2001 Jun 26 | 7.788 | 0.71 ± 0.12 | 8.040 | 0.45 ± 0.08 | 8.140 | 0.91 ± 0.09 | |
| **Hylonome** | | | | 0.77 ± 0.08 | | 0.38 ± 0.06 | | 0.91 ± 0.09 | |
| **1995 $SM_{55}$** | 20.61 ± 0.02 | 2000 Dec 21 | 6.820 | 0.68 ± 0.04 | 7.013 | 0.39 ± 0.04 | 7.080 | 0.79 ± 0.04 | |
| **1995 $SM_{55}$** | | 2000 Dec 21 | 7.221 | 0.62 ± 0.05 | 7.396 | 0.41 ± 0.04 | 7.457 | 0.77 ± 0.04 | |
| **1995 $SM_{55}$** | | | | 0.66 ± 0.03 | | 0.40 ± 0.03 | | 0.78 ± 0.03 | |
| **1995 $TL_{8}$** | 21.71 ± 0.05 | 2000 Dec 23 | 6.691 | 0.60 ± 0.08 | 7.035 | 0.38 ± 0.07 | 7.129 | 0.67 ± 0.07 | |
| **1995 $TL_{8}$** | | 2000 Dec 23 | 7.413 | 0.72 ± 0.06 | 7.738 | 0.41 ± 0.05 | 7.852 | 0.67 ± 0.07 | |
| **1995 $TL_{8}$** | | | | 0.68 ± 0.05 | | 0.40 ± 0.05 | | 0.67 ± 0.05 | |
| **1996 $TK_{66}$** | 22.95 ± 0.05 | 2001 Aug 14 | 8.801 | 0.95 ± 0.08 | 12.829 | 0.81 ± 0.05 | 13.451 | 1.42 ± 0.06 | |
| **1996 $TK_{66}$** | | | | 0.95 ± 0.08 | | 0.81 ± 0.05 | | 1.42 ± 0.06 | |
| **1998 $BU_{48}$** | 21.53 ± 0.03 | 2000 Dec 24 | 11.684 | 0.98 ± 0.06 | 12.008 | 0.62 ± 0.04 | 12.086 | 1.14 ± 0.04 | |
| **1998 $BU_{48}$** | | 2000 Dec 24 | 12.356 | 1.02 ± 0.04 | 12.679 | 0.65 ± 0.04 | 12.770 | 1.22 ± 0.05 | |
| **1998 $BU_{48}$** | | | | 1.01 ± 0.04 | | 0.64 ± 0.03 | | 1.17 ± 0.04 | |
| **1998 $HK_{151}$** | 22.24 ± 0.03 | 2001 Jun 26 | 9.216 | 0.72 ± 0.05 | 8.756 | 0.52 ± 0.04 | 9.598 | 0.88 ± 0.05 | |
| **1998 $HK_{151}$** | | | | 0.72 ± 0.05 | | 0.52 ± 0.04 | | 0.88 ± 0.05 | YES* |
| **1998 $QM_{107}$** | 23.22 ± 0.06 | 2001 Aug 12 | | | 9.635 | 0.68 ± 0.08 | | | |
| **1998 $QM_{107}$** | | 2001 Aug 12 | | | 11.330 | 0.67 ± 0.08 | | | |
| **1998 $QM_{107}$** | | | | | | 0.68 ± 0.06 | | | |
| **1998 $TF_{35}$** | 22.11 ± 0.04 | 2000 dec 21 | 7.995 | 1.17 ± 0.06 | 8.319 | 0.70 ± 0.06 | 8.445 | 1.40 ± 0.06 | |
| **1998 $TF_{35}$** | | 2000 dec 21 | 8.830 | 1.18 ± 0.06 | 9.413 | 0.72 ± 0.05 | 9.554 | 1.41 ± 0.06 | |
| **1998 $TF_{35}$** | | | | 1.17 ± 0.05 | | 0.71 ± 0.04 | | 1.41 ± 0.04 | |
| **1998 $VG_{44}$** | 21.56 ± 0.04 | 2000 Dec 23 | 9.233 | 0.89 ± 0.06 | 9.475 | 0.58 ± 0.05 | 9.551 | 1.10 ± 0.07 | |
| **1998 $VG_{44}$** | | 2000 Dec 23 | 9.751 | 0.88 ± 0.07 | 9.992 | 0.59 ± 0.05 | 10.067 | 1.08 ± 0.06 | |
| **1998 $VG_{44}$** | | | | 0.89 ± 0.05 | | 0.58 ± 0.04 | | 1.08 ± 0.05 | |
| **1998 $WH_{24}$** | 21.31 ± 0.05 | 2000 Dec 22 | 8.357 | 0.94 ± 0.05 | 8.596 | 0.63 ± 0.04 | 8.675 | 1.26 ± 0.05 | |
| **1998 $WH_{24}$** | | 2000 Dec 22 | 8.870 | 0.95 ± 0.04 | 9.434 | 0.63 ± 0.04 | 9.244 | 1.24 ± 0.05 | |
| **1998 $WH_{24}$** | | | | 0.95 ± 0.03 | | 0.63 ± 0.03 | | 1.25 ± 0.04 | |
| **1998 $WW_{24}$** | 23.43 ± 0.05 | 2000 Dec 24 | 9.297 | 0.69 ± 0.08 | 9.687 | 0.42 ± 0.06 | 9.878 | 1.15 ± 0.07 | |
| **1998 $WW_{24}$** | | | | 0.69 ± 0.08 | | 0.42 ± 0.06 | | 1.15 ± 0.07 | |
| **1999 $CC_{158}$** | 22.16 ± 0.03 | 2000 Dec 22 | 12.021 | 1.02 ± 0.06 | 12.345 | 0.64 ± 0.05 | 12.439 | 1.29 ± 0.06 | |
| **1999 $CC_{158}$** | | 2000 Dec 22 | 12.814 | 1.01 ± 0.06 | 13.141 | 0.63 ± 0.05 | 13.234 | 1.27 ± 0.06 | |
| **1999 $CC_{158}$** | | | | 1.01 ± 0.04 | | 0.63 ± 0.03 | | 1.28 ± 0.04 | |
| **1999 $CD_{158}$** | 22.09 ± 0.03 | 2000 Dec 22 | 10.619 | 0.86 ± 0.06 | 10.946 | 0.55 ± 0.05 | 11.044 | 1.12 ± 0.06 | |
| **1999 $CD_{158}$** | | 2000 Dec 22 | 11.315 | 0.86 ± 0.06 | 11.637 | 0.53 ± 0.05 | 11.751 | 1.14 ± 0.06 | |
| **1999 $CD_{158}$** | | | | 0.86 ± 0.04 | | 0.54 ± 0.03 | | 1.13 ± 0.04 | |
| **1999 $CL_{158}$** | 22.32 ± 0.05 | 2000 Dec 23 | 11.667 | 0.72 ± 0.08 | 12.158 | 0.43 ± 0.07 | 12.283 | 0.85 ± 0.08 | |
| **1999 $CL_{158}$** | | 2000 Dec 23 | 12.641 | 0.84 ± 0.06 | 13.133 | 0.38 ± 0.04 | 13.250 | 0.88 ± 0.09 | |
| **1999 $CL_{158}$** | | | | 0.80 ± 0.06 | | 0.39 ± 0.04 | | 0.86 ± 0.06 | |
| **1999 $DE_{9}$** | 20.67 ± 0.03 | 2000 Dec 22 | 15.085 | 0.94 ± 0.04 | 15.260 | 0.60 ± 0.04 | 15.323 | 1.20 ± 0.05 | |
| **1999 $DE_{9}$** | | 2000 Dec 22 | 15.494 | 0.94 ± 0.05 | 15.668 | 0.59 ± 0.04 | 15.728 | 1.15 ± 0.05 | |
| **1999 $DE_{9}$** | | | | 0.94 ± 0.03 | | 0.59 ± 0.03 | | 1.17 ± 0.03 | |
| **1999 $DF_{9}$** | 22.48 ± 0.06 | 2000 Dec 23 | 14.533 | 0.92 ± 0.07 | 14.854 | 0.70 ± 0.07 | 14.946 | 1.30 ± 0.06 | |
| **1999 $DF_{9}$** | | 2000 Dec 23 | 15.203 | 0.91 ± 0.07 | 15.517 | 0.72 ± 0.06 | 15.617 | 1.49 ± 0.06 | |
| **1999 $DF_{9}$** | | | | 0.92 ± 0.06 | | 0.71 ± 0.05 | | 1.36 ± 0.06 | V-I |
| **1999 $OX_{3}$** | 22.15 ± 0.05 | 2001 Jun 27 | 11.365 | 1.21 ± 0.07 | 11.019 | 0.62 ± 0.05 | 11.595 | 1.44 ± 0.04 | |
| **1999 $OX_{3}$** | | 2001 Jun 27 | | | | | 11.856 | 1.42 ± 0.04 | |
| **1999 $OX_{3}$** | | | | 1.21 ± 0.07 | | 0.62 ± 0.05 | | 1.43 ± 0.05 | |
| **1999 $OY_{3}$** | 22.47 ± 0.03 | 2001 Aug 14 | 8.801 | 0.78 ± 0.05 | 8.554 | 0.21 ± 0.04 | 9.177 | 0.47 ± 0.05 | |
| **1999 $OY_{3}$** | | 2001 Aug 14 | 9.771 | 0.72 ± 0.05 | 9.523 | 0.31 ± 0.04 | 10.147 | 0.67 ± 0.05 | |
| **1999 OY$_3$** | | | | 0.75 ± 0.03 | | 0.26 ± 0.03 | | 0.57 ± 0.04 | YES |
| **1999 TD$_{10}$** | 20.09 ± 0.02 | 2000 Dec 22 | 6.329 | 0.75 ± 0.03 | 6.502 | 0.52 ± 0.03 | 6.560 | 0.95 ± 0.04 | |
| **1999 TD$_{10}$** | | 2000 Dec 22 | 6.699 | 0.70 ± 0.03 | 6.901 | 0.49 ± 0.04 | 6.981 | 0.99 ± 0.04 | |
| **1999 TD$_{10}$** | | | | 0.72 ± 0.02 | | 0.51 ± 0.02 | | 0.97 ± 0.03 | |
| **1999 UG$_5$** | 19.73 ± 0.01 | 2000 Dec 22 | 7.655 | 1.08 ± 0.03 | 7.815 | 0.62 ± 0.03 | 7.878 | 1.32 ± 0.04 | |
| **1999 UG$_5$** | | 2000 Dec 22 | 8.002 | 1.04 ± 0.03 | 8.164 | 0.65 ± 0.03 | 8.222 | 1.26 ± 0.04 | |
| **1999 UG$_5$** | | | | 1.05 ± 0.02 | | 0.64 ± 0.02 | | 1.29 ± 0.03 | |
| **1999 XX$_{143}$** | 23.00 ± 0.05 | 2000 Dec 21 | 12.591 | 0.99 ± 0.07 | 12.914 | 0.61 ± 0.06 | 13.046 | 0.99 ± 0.06 | |
| **1999 XX$_{143}$** | | 2000 Dec 21 | 13.604 | 1.09 ± 0.09 | 13.934 | 0.58 ± 0.05 | 14.074 | 1.04 ± 0.06 | |
| **1999 XX$_{143}$** | | | | 1.02 ± 0.06 | | 0.59 ± 0.04 | | 1.01 ± 0.04 | |
| **2000 FE$_8$** | 22.59 ± 0.05 | 2000 Dec 24 | 14.207 | 0.74 ± 0.07 | 14.533 | 0.50 ± 0.05 | 14.624 | 0.98 ± 0.05 | |
| **2000 FE$_8$** | | 2000 Dec 24 | 14.881 | 0.77 ± 0.09 | 15.205 | 0.45 ± 0.07 | 15.284 | 0.99 ± 0.08 | |
| **2000 FE$_8$** | | | | 0.75 ± 0.06 | | 0.48 ± 0.04 | | 0.99 ± 0.05 | |
| **2000 GP$_{183}$** | 21.96 ± 0.05 | 2001 Jun 27 | 6.910 | 0.78 ± 0.05 | 7.430 | 0.36 ± 0.04 | 7.581 | 0.87 ± 0.04 | |
| **2000 GP$_{183}$** | | 2001 Jun 27 | 7.985 | 0.75 ± 0.06 | 8.498 | 0.42 ± 0.05 | 8.625 | 0.74 ± 0.06 | |
| **2000 GP$_{183}$** | | | | 0.77 ± 0.04 | | 0.39 ± 0.04 | | 0.82 ± 0.04 | V-I |
| **2000 OJ$_{67}$** | 22.93 ± 0.07 | 2001 Jun 27 | 12.676 | 1.10 ± 0.08 | 13.229 | 0.69 ± 0.06 | 13.354 | 1.27 ± 0.07 | |
| **2000 OJ$_{67}$** | | 2001 Jun 27 | 13.978 | 0.99 ± 0.09 | 14.487 | 0.65 ± 0.07 | | | |
| **2000 OJ$_{67}$** | | | | 1.05 ± 0.06 | | 0.67 ± 0.05 | | 1.27 ± 0.07 | |
| **2000 OK$_{67}$** | 22.81 ± 0.07 | 2001 Jun 26 | 13.489 | 1.02 ± 0.10 | 13.086 | 0.64 ± 0.07 | | | |
| **2000 OK$_{67}$** | | 2001 Jun 26 | 14.227 | 0.78 ± 0.10 | 14.620 | 0.68 ± 0.08 | 13.950 | 1.22 ± 0.08 | |
| **2000 OK$_{67}$** | | | | 0.89 ± 0.08 | | 0.65 ± 0.05 | | 1.22 ± 0.08 | B-V |
| **2000 QC$_{243}$** | 20.31 ± 0.03 | 2001 Aug 14 | 11.970 | 0.66 ± 0.04 | 11.878 | 0.43 ± 0.04 | 12.049 | 0.90 ± 0.04 | |
| **2000 QC$_{243}$** | | 2001 Aug 14 | 12.244 | 0.69 ± 0.04 | 12.151 | 0.45 ± 0.04 | 12.323 | 0.92 ± 0.04 | |
| **2000 QC$_{243}$** | | | | 0.67 ± 0.03 | | 0.44 ± 0.03 | | 0.91 ± 0.03 | |
| **2000 WR$_{106}$** | 20.34 ± 0.02 | 2000 Dec 21 | 10.898 | 0.94 ± 0.03 | 11.074 | 0.62 ± 0.02 | 11.134 | 1.23 ± 0.02 | |
| **2000 WR$_{106}$** | | 2000 Dec 21 | 11.289 | 1.02 ± 0.03 | 11.464 | 0.61 ± 0.02 | 11.518 | 1.21 ± 0.02 | |
| **2000 WR$_{106}$** | | | | 0.92 ± 0.03 | | 0.61 ± 0.02 | | 1.22 ± 0.02 | |
| **2001 KA$_{77}$** | 22.66 ± 0.05 | 2001 Aug 14 | 6.710 | 1.00 ± 0.77 | 6.528 | 0.76 ± 0.55 | 6.934 | 1.59 ± 0.06 | |
| **2001 KA$_{77}$** | | 2001 Aug 14 | 7.362 | 1.10 ± 0.78 | 7.181 | 0.57 ± 0.61 | 7.586 | 1.37 ± 0.06 | |
| **2001 KA$_{77}$** | | | | 1.05 ± 0.05 | | 0.67 ± 0.04 | | 1.48 ± 0.04 | YES |
| **2001 KD$_{77}$** | 21.97 ± 0.02 | 2001 Aug 12 | 7.480 | 1.12 ± 0.06 | 7.366 | 0.67 ± 0.05 | 7.587 | 1.23 ± 0.07 | |
| **2001 KD$_{77}$** | | 2001 Aug 12 | 7.831 | | 7.716 | 0.62 ± 0.06 | 7.937 | 1.20 ± 0.07 | |
| **2001 KD$_{77}$** | | | | 1.12 ± 0.06 | | 0.65 ± 0.04 | | 1.22 ± 0.05 | |
| **2001 KX$_{76}$** | 20.39 ± 0.02 | 2001 Aug 12 | 6.487 | 1.03 ± 0.04 | 6.343 | 0.61 ± 0.04 | 6.597 | 1.18 ± 0.05 | |
| **2001 KX$_{76}$** | | 2001 Aug 12 | 6.887 | 1.03 ± 0.03 | 6.759 | 0.60 ± 0.04 | 7.001 | 1.21 ± 0.05 | |
| **2001 KX$_{76}$** | | | | 1.03 ± 0.03 | | 0.61 ± 0.03 | | 1.19 ± 0.04 | |

Notes: Individual B-V, V-R, V-I color index are indicated as well as their 1σ error. For each object, the shaded line indicates the weighted mean of the individual measurements. [1]UT start time in hour of the corresponding B, R, or I exposure. [2]This column indicates whether a true (no rotation effect) color variation is detected. [*]color different from previous measurement (Doressoundiram et al. 2001)

## Table 4: Mean Colors and estimated size of Centaurs and TNOs.

| *Object* | *V* | *B-V* | *V-R* | *R-I* | *V-I* | $H_V$[1] | *Size* |
|---|---|---|---|---|---|---|---|
| **Solar colors** | | 0.67 | 0.36 | 0.33 | 0.69 | | |
| **10370 Hylonome** | 22.59 ± 0.08 | 0.77 ± 0.08 | 0.38 ± 0.06 | 0.53 ± 0.09 | 0.91 ± 0.09 | 9.49 ± 0.08 | 84 |
| **1995 $SM_{55}$** | 20.61 ± 0.02 | 0.66 ± 0.03 | 0.40 ± 0.03 | 0.38 ± 0.03 | 0.78 ± 0.03 | 4.53 ± 0.02 | 826 |
| **1995 $TL_8$** | 21.71 ± 0.05 | 0.68 ± 0.05 | 0.40 ± 0.05 | 0.27 ± 0.05 | 0.67 ± 0.05 | 5.31 ± 0.05 | 576 |
| **1996 $TK_{66}$** | 22.95 ± 0.05 | 0.95 ± 0.08 | 0.81 ± 0.05 | 0.61 ± 0.06 | 1.42 ± 0.06 | 6.53 ± 0.05 | 328 |
| **1998 $BU_{48}$** | 21.53 ± 0.03 | 1.01 ± 0.04 | 0.64 ± 0.03 | 0.53 ± 0.04 | 1.17 ± 0.04 | 7.00 ± 0.03 | 265 |
| **1998 $HK_{151}$** | 22.24 ± 0.03 | 0.72 ± 0.05 | 0.52 ± 0.04 | 0.36 ± 0.05 | 0.88 ± 0.05 | 7.32 ± 0.03 | 228 |
| **1998 $QM_{107}$** | 23.22 ± 0.06 | | 0.68 ± 0.06 | | | 10.76 ± 0.06 | 47 |
| **1998 $TF_{35}$** | 22.11 ± 0.04 | 1.17 ± 0.05 | 0.71 ± 0.04 | 0.70 ± 0.04 | 1.41 ± 0.04 | 9.17 ± 0.04 | 97 |
| **1998 $VG_{44}$** | 21.56 ± 0.04 | 0.89 ± 0.05 | 0.58 ± 0.04 | 0.50 ± 0.05 | 1.08 ± 0.05 | 6.68 ± 0.04 | 306 |
| **1998 $WH_{24}$** | 21.31 ± 0.05 | 0.95 ± 0.03 | 0.63 ± 0.03 | 0.62 ± 0.04 | 1.25 ± 0.04 | 5.00 ± 0.05 | 665 |
| **1998 $WW_{24}$** | 23.43 ± 0.05 | 0.69 ± 0.08 | 0.42 ± 0.06 | 0.73 ± 0.07 | 1.15 ± 0.07 | 8.35 ± 0.05 | 142 |
| **1999 $CC_{158}$** | 22.16 ± 0.03 | 1.01 ± 0.04 | 0.63 ± 0.03 | 0.65 ± 0.04 | 1.28 ± 0.04 | 5.81 ± 0.03 | 458 |
| **1999 $CD_{158}$** | 22.09 ± 0.03 | 0.86 ± 0.04 | 0.54 ± 0.03 | 0.59 ± 0.04 | 1.13 ± 0.04 | 5.21 ± 0.03 | 603 |
| **1999 $CL_{158}$** | 22.32 ± 0.05 | 0.80 ± 0.06 | 0.39 ± 0.04 | 0.47 ± 0.06 | 0.86 ± 0.06 | 6.92 ± 0.05 | 275 |
| **1999 $DE_9$** | 20.67 ± 0.03 | 0.94 ± 0.03 | 0.59 ± 0.03 | 0.58 ± 0.03 | 1.17 ± 0.03 | 5.22 ± 0.03 | 602 |
| **1999 $DF_9$** | 22.48 ± 0.06 | 0.92 ± 0.06 | 0.71 ± 0.05 | 0.65 ± 0.06 | 1.36 ± 0.06 | 6.35 ± 0.06 | 357 |
| **1999 $OX_3$** | 22.15 ± 0.05 | 1.21 ± 0.07 | 0.62 ± 0.05 | 0.81 ± 0.05 | 1.43 ± 0.05 | 7.77 ± 0.05 | 185 |
| **1999 $OY_3$** | 22.47 ± 0.03 | 0.75 ± 0.03 | 0.26 ± 0.03 | 0.31 ± 0.04 | 0.57 ± 0.04 | 6.54 ± 0.03 | 326 |
| **1999 $TD_{10}$** | 20.09 ± 0.02 | 0.72 ± 0.02 | 0.51 ± 0.02 | 0.55 ± 0.03 | 0.97 ± 0.03 | 8.76 ± 0.02 | 118 |
| **1999 $UG_5$** | 19.73 ± 0.01 | 1.05 ± 0.02 | 0.64 ± 0.02 | 0.65 ± 0.03 | 1.29 ± 0.03 | 10.52 ± 0.01 | 52 |
| **1999 $XX_{143}$** | 23.00 ± 0.05 | 1.02 ± 0.06 | 0.59 ± 0.04 | 0.42 ± 0.04 | 1.01 ± 0.04 | 9.01 ± 0.05 | 105 |
| **2000 $FE_8$** | 22.59 ± 0.05 | 0.75 ± 0.06 | 0.48 ± 0.04 | 0.50 ± 0.05 | 0.98 ± 0.05 | 6.90 ± 0.05 | 277 |
| **2000 $GP_{183}$** | 21.96 ± 0.05 | 0.77 ± 0.04 | 0.39 ± 0.04 | 0.43 ± 0.04 | 0.82 ± 0.04 | 6.09 ± 0.05 | 402 |
| **2000 $OJ_{67}$** | 22.93 ± 0.07 | 1.05 ± 0.06 | 0.67 ± 0.05 | 0.60 ± 0.07 | 1.27 ± 0.07 | 6.55 ± 0.07 | 325 |
| **2000 $OK_{67}$** | 22.81 ± 0.07 | 0.89 ± 0.08 | 0.65 ± 0.05 | 0.57 ± 0.08 | 1.22 ± 0.08 | 6.59 ± 0.07 | 320 |
| **2000 $QC_{243}$** | 20.31 ± 0.03 | 0.67 ± 0.03 | 0.44 ± 0.03 | 0.47 ± 0.03 | 0.91 ± 0.03 | 7.45 ± 0.03 | 215 |
| **2000 $WR_{106}$ (Varuna)** | 20.34 ± 0.02 | 0.92 ± 0.03 | 0.61 ± 0.02 | 0.61 ± 0.02 | 1.22 ± 0.02 | 4.02 ± 0.02 | 788 |
| **2001 $KA_{77}$** | 22.66 ± 0.05 | 1.05 ± 0.05 | 0.67 ± 0.04 | 0.81 ± 0.04 | 1.48 ± 0.04 | 5.64 ± 0.05 | 494 |
| **2001 $KD_{77}$** | 21.97 ± 0.02 | 1.12 ± 0.03 | 0.65 ± 0.04 | 0.57 ± 0.05 | 1.22 ± 0.05 | 6.34 ± 0.02 | 358 |
| **2001 $KX_{76}$** | 20.39 ± 0.02 | 1.03 ± 0.03 | 0.61 ± 0.03 | 0.58 ± 0.04 | 1.19 ± 0.04 | 3.89 ± 0.02 | 1110 |

Notes: [1]$H_V$=absolute V magnitude. The observed V magnitude has been averaged over several measurements. When multiple colors were available, a weighted mean has been computed. The equivalent diameter has been derived from the absolute magnitude and assuming an albedo of 0.04. (except for 20000 Varuna for which an albedo of 0.07 has been taken following Jewitt, D. C., Aussel, H., & Evans (2001).

**Table 5: All other data published by our group.**

| *Object* | *group* | *V* | *B-V* | *V-R* | *R-I* | $H_V$[1] | *Size (km)* | *ref* |
|---|---|---|---|---|---|---|---|---|
| **Solar colors** | | | 0.67 | 0.36 | 0.33 | | | |
| **1993 FW** | Classical | 23.47 ± 0.02 | 1.01 ± 0.09 | 0.66 ± 0.04 | 0.39 ± 0.09 | 7.15 ± 0.02 | 247 | 2 |
| **1994 JR$_1$** | Plutino | 23.30 ± 0.30 | 1.01 ± 0.18 | 0.60 ± 0.12 | 1.12 ± 0.12 | 7.70 ± 0.30 | 164 | 1 |
| **1994 TB** | Plutino | 22.00 ± 0.30 | 1.10 ± 0.18 | 0.74 ± 0.11 | 0.72 ± 0.08 | 7.10 ± 0.30 | 216 | 1 |
| **1995 HM$_5$** | Plutino | 23.40 ± 0.02 | 0.67 ± 0.06 | 0.55 ± 0.04 | 0.37 ± 0.11 | 8.23 ± 0.02 | 150 | 2 |
| **1995 QY$_9$** | Plutino | 22.20 ± 0.30 | 0.74 ± 0.20 | 0.47 ± 0.12 | | 7.50 ± 0.30 | 180 | 1 |
| **1996 TL$_{66}$** | Scattered | 21.00 ± 0.30 | 0.74 ± 0.08 | 0.44 ± 0.05 | 0.26 ± 0.07 | 5.40 ± 0.30 | 473 | 1 |
| **1996 TO$_{66}$** | Classical | 21.20 ± 0.30 | 0.72 ± 0.07 | 0.40 ± 0.04 | 0.38 ± 0.04 | 4.50 ± 0.30 | 716 | 1 |
| **1996 TP$_{66}$** | Plutino | 21.20 ± 0.30 | 1.14 ± 0.14 | 0.65 ± 0.13 | 0.43 ± 0.14 | 6.80 ± 0.30 | 248 | 1 |
| **1997 CQ$_{29}$** | Classical | 23.53 ± 0.02 | 0.99 ± 0.12 | 0.68 ± 0.06 | 0.62 ± 0.09 | 7.38 ± 0.03 | 222 | 2 |
| **1997 CR$_{29}$** | Classical | 23.71 ± 0.08 | 0.79 ± 0.10 | 0.47 ± 0.08 | 0.68 ± 0.12 | 7.43 ± 0.08 | 217 | 3 |
| **1997 CS$_{29}$** | Classical | 21.97 ± 0.01 | 1.05 ± 0.06 | 0.66 ± 0.02 | 0.53 ± 0.04 | 5.54 ± 0.01 | 518 | 2 |
| **1997 CU$_{29}$** | Classical | 23.25 ± 0.06 | 1.05 ± 0.10 | 0.66 ± 0.06 | 0.56 ± 0.06 | 6.77 ± 0.06 | 294 | 3 |
| **1998 FS$_{144}$** | Classical | 23.39 ± 0.03 | 0.91 ± 0.08 | 0.56 ± 0.07 | | 7.14 ± 0.03 | 248 | 2 |
| **1998 KG$_{62}$** | Classical | 23.55 ± 0.05 | 1.00 ± 0.06 | 0.53 ± 0.05 | 0.64 ± 0.04 | 6.90 ± 0.05 | 302 | 3 |
| **1998 SG$_{35}$** | Centaur | 21.43 ± 0.06 | 0.89 ± 0.11 | 0.43 ± 0.08 | 0.59 ± 0.08 | 11.16 ± 0.06 | 39 | 3 |
| **1998 SN$_{165}$** | Classical | 21.55 ± 0.06 | 0.82 ± 0.08 | 0.33 ± 0.08 | 0.51 ± 0.08 | 5.67 ± 0.06 | 488 | 3 |
| **1999 HB$_{12}$** | Scattered | 22.67 ± 0.04 | 0.87 ± 0.06 | 0.50 ± 0.05 | 0.32 ± 0.08 | 7.10 ± 0.06 | 272 | 3 |
| **1999 HR$_{11}$** | Classical | 23.94 ± 0.05 | 0.92 ± 0.12 | 0.53 ± 0.10 | 0.80 ± 0.07 | 7.55 ± 0.05 | 224 | 3 |
| **1999 HS$_{11}$** | Classical | 23.41 ± 0.05 | 1.01 ± 0.16 | 0.68 ± 0.10 | 0.60 ± 0.09 | 6.98 ± 0.05 | 312 | 3 |
| **1999 RY$_{215}$** | Classical | 22.99 ± 0.06 | 0.80 ± 0.10 | 0.48 ± 0.05 | 0.78 ± 0.08 | 7.22 ± 0.06 | 240 | 3 |
| **1999 TC$_{36}$** | Plutinos | 20.49 ± 0.05 | 0.99 ± 0.09 | 0.65 ± 0.06 | 0.72 ± 0.07 | 5.40 ± 0.05 | 552 | 3 |
| **2000 EB$_{173}$** | Plutinos | 20.00 ± 0.01 | 0.96 ± 0.02 | 0.53 ± 0.01 | 0.66 ± 0.01 | 5.09 ± 0.01 | 696 | 3 |
| **2000 PE$_{30}$** | Scattered | 22.04 ± 0.04 | 0.71 ± 0.05 | 0.38 ± 0.04 | 0.45 ± 0.04 | 6.16 ± 0.04 | 390 | 3 |

Notes: [1]$H_V$=absolute V magnitude. Only objects with available B-V and V-R colors are reported in this table. [1] Barucci, M. A., A. Doressoundiram, D. J. Tholen, M. Fulchignoni, and M. Lazzarin, 1999. Icarus, 142, 476. [2] Barucci, M. A., Romon, J., Doressoundiram, A. and D. J. Tholen, 2000. AJ, 120, 496. [3] Doressoundiram A., Barucci, M. A., Romon J, and Veillet, Ch. 2001. Icarus, 154, 277.

## Table 6: Correlations.

| *Quantity*[1] | *Quantity*[1] | *group*[2] | $r_{corr}$[3] | $P(r>r_{corr})$[4] | *Significance*[5] |
|---|---|---|---|---|---|
| **B-V** | V-R | Our whole dataset (N=52) | **0.79** | **$5\ 10^{-12}$** | **>8σ** |
| **B-V** | R-I | Our whole dataset (N=51) | **0.47** | **0.0006** | **3.4σ** |
| **V-R** | R-I | Our whole dataset (N=51) | **0.47** | **0.0006** | **3.4σ** |
| **B-R** | a | Centaurs (N=8) | 0.45 | 0.26 | 1.1σ |
| **B-R** | e | Centaurs (N=8) | 0.62 | 0.10 | 1.6σ |
| **B-R** | i | Centaurs (N=8) | -0.43 | 0.29 | 1.1σ |
| **B-R** | a | Classical (N=22) | 0.22 | 0.33 | 1.0σ |
| **B-R** | e | Classical (N=22) | **-0.60** | **0.003** | **3.0σ** |
| **B-V** | i | Classical (N=22) | **-0.80** | **$9\ 10^{-6}$** | **4.4σ** |
| **B-R** | i | Classical (N=22) | **-0.72** | **0.0002** | **3.8σ** |
| **B-I** | i | Classical (N=21) | **-0.69** | **0.0006** | **3.5σ** |
| **R-I** | i | Classical (N=21) | -0.19 | | |
| **B-R** | q | Classical (N=22) | **0.76** | **$4\ 10^{-5}$** | **4.1σ** |
| **B-R** | H | Classical (N=22) | ~0 | | |
| **B-R** | size | Classical (N=22) | ~0 | | |
| **B-R** | $V_{rms}$ | Classical (N=22) | **-0.77** | **$3\ 10^{-5}$** | **4.2σ** |
| **B-R** | e | Classical+Scattered (N=30) | **-0.60** | **0.0005** | **3.5σ** |
| **B-R** | i | Classical+Scattered (N=30) | **-0.53** | **0.003** | **3.0σ** |
| **B-R** | q | Classical+Scattered (N=30) | **0.68** | **$4\ 10^{-5}$** | **4.1σ** |
| **B-R** | e | All published Classical (N=50) | -0.21 | 0.15 | 1.5σ |
| **B-R** | i | All published Classical (N=50) | **-0.53** | **$8\ 10^{-5}$** | **4.0σ** |
| **B-R** | q | All published Classical (N=50) | -0.25 | 0.08 | 1.8σ |
| **B-R** | $V_{rms}$ | All published Classical (N=50) | **-0.49** | **0.0004** | **3.6σ** |
| **B-R** | e | Plutinos (N=13) | 0.31 | 0.30 | 1.0σ |
| **B-R** | i | Plutinos (N=13) | 0.06 | | |
| **B-R** | e | All published Plutinos (N=25) | -0.03 | | |
| **B-R** | i | All published Plutinos (N=25) | 0.15 | | |

[1]*a*=semimajor axis, *e*=eccentricity, *i*=inclination, q=perihelion, H=absolute magnitude, size=estimated diameter, $V_{rms}$= mean excitation velocity.
[2]Dynamical class of TNOs and number of measurements in the sample.
[3]Spearman's rank correlation statistic
[4]Probability of obtaining a higher or equal coefficient from a uncorrelated set of data. P < 0.003 indicates a correlation with >3σ significance.
[5]Significance of the correlation, derived from the confidence level (1-P) and assuming Gaussian statistics.
99.0% confidence level means a 2.6σ significance.
99.9% confidence level means a 3.3σ significance.
All the strong and significant correlations are marked as bold.

**Table 7: Orbital and physical characteristics of the dynamically Cold and Hot Classical TNOs.**

| *Object* | *a*[1] *(AU)* | *e*[2] | *i*[3] *(deg)* | *$H_V$*[4] *(mag)* | *Size*[5] *(km)* | *B-R*[6] *(mag)* |
|---|---|---|---|---|---|---|
| **Dynamically Cold Classical TNOs** | | | | | | |
| **1998 $KG_{62}$** | 43.206 | 0.055 | 0.8 | 6.90 ± 0.05 | 277 | 1.53 ± 0.06 |
| **2000 $OJ_{67}$** | 42.863 | 0.098 | 1.1 | 6.55 ± 0.07 | 325 | 1.72 ± 0.06 |
| **1997 $CU_{29}$** | 43.763 | 0.032 | 1.5 | 6.77 ± 0.06 | 294 | 1.71 ± 0.10 |
| **1997 $CS_{29}$** | 44.194 | 0.016 | 2.2 | 5.54 ± 0.01 | 518 | 1.71 ± 0.06 |
| **1999 $HS_{11}$** | 44.071 | 0.017 | 2.6 | 6.98 ± 0.05 | 267 | 1.69 ± 0.16 |
| **1997 $CQ_{29}$** | 45.415 | 0.124 | 2.9 | 7.38 ± 0.03 | 222 | 1.67 ± 0.12 |
| **1999 $HR_{11}$** | 43.752 | 0.038 | 3.3 | 7.55 ± 0.05 | 205 | 1.45 ± 0.12 |
| **1996 $TK_{66}$** | 42.772 | 0.013 | 3.3 | 6.53 ± 0.05 | 328 | 1.76 ± 0.08 |
| **2000 $OK_{67}$** | 46.432 | 0.141 | 4.9 | 6.59 ± 0.07 | 319 | 1.54 ± 0.08 |
| **Dynamically Hot Classical TNOs** | | | | | | |
| **1993 FW** | 43.731 | 0.047 | 7.8 | 7.15 ± 0.02 | 246 | 1.67 ± 0.09 |
| **1999 $DF_9$** | 46.678 | 0.148 | 9.8 | 6.35 ± 0.06 | 356 | 1.63 ± 0.06 |
| **1998 $FS_{144}$** | 41.913 | 0.023 | 9.9 | 7.14 ± 0.03 | 248 | 1.47 ± 0.08 |
| **1999 $CL_{158}$** | 41.830 | 0.215 | 10.0 | 6.92 ± 0.05 | 274 | 1.19 ± 0.06 |
| **2001 $KA_{77}$** | 47.026 | 0.070 | 12.0 | 5.64 ± 0.05 | 495 | 1.72 ± 0.05 |
| **1998 $WH_{24}$** | 46.079 | 0.110 | 12.0 | 5.00 ± 0.05 | 664 | 1.58 ± 0.03 |
| **2000 $WR_{106}$** | 43.290 | 0.054 | 17.1 | 4.02 ± 0.02 | 1043 | 1.53 ± 0.03 |
| **1997 $CR_{29}$** | 47.358 | 0.216 | 19.1 | 7.43 ± 0.08 | 217 | 1.26 ± 0.10 |
| **1999 $RY_{215}$** | 45.144 | 0.236 | 22.2 | 7.22 ± 0.06 | 239 | 1.28 ± 0.10 |
| **1999 $OY_3$** | 43.606 | 0.166 | 24.3 | 6.54 ± 0.03 | 327 | 1.01 ± 0.03 |
| **1999 $CD_{158}$** | 43.922 | 0.138 | 25.4 | 5.21 ± 0.03 | 603 | 1.40 ± 0.04 |
| **1995 $SM_{55}$** | 41.991 | 0.109 | 27.0 | 4.53 ± 0.02 | 825 | 1.06 ± 0.03 |
| **1996 $TO_{66}$** | 43.300 | 0.116 | 27.5 | 4.50 ± 0.30 | 836 | 1.12 ± 0.07 |

[1]*a*=semimajor axis, [2]*e*=eccentricity, [3]*i*=inclination, [4]$H_V$=absolute V magnitude, [5]size=estimated diameter. [6]B-R= color index. B-R=1.03 for the Sun.

**Figure 1**

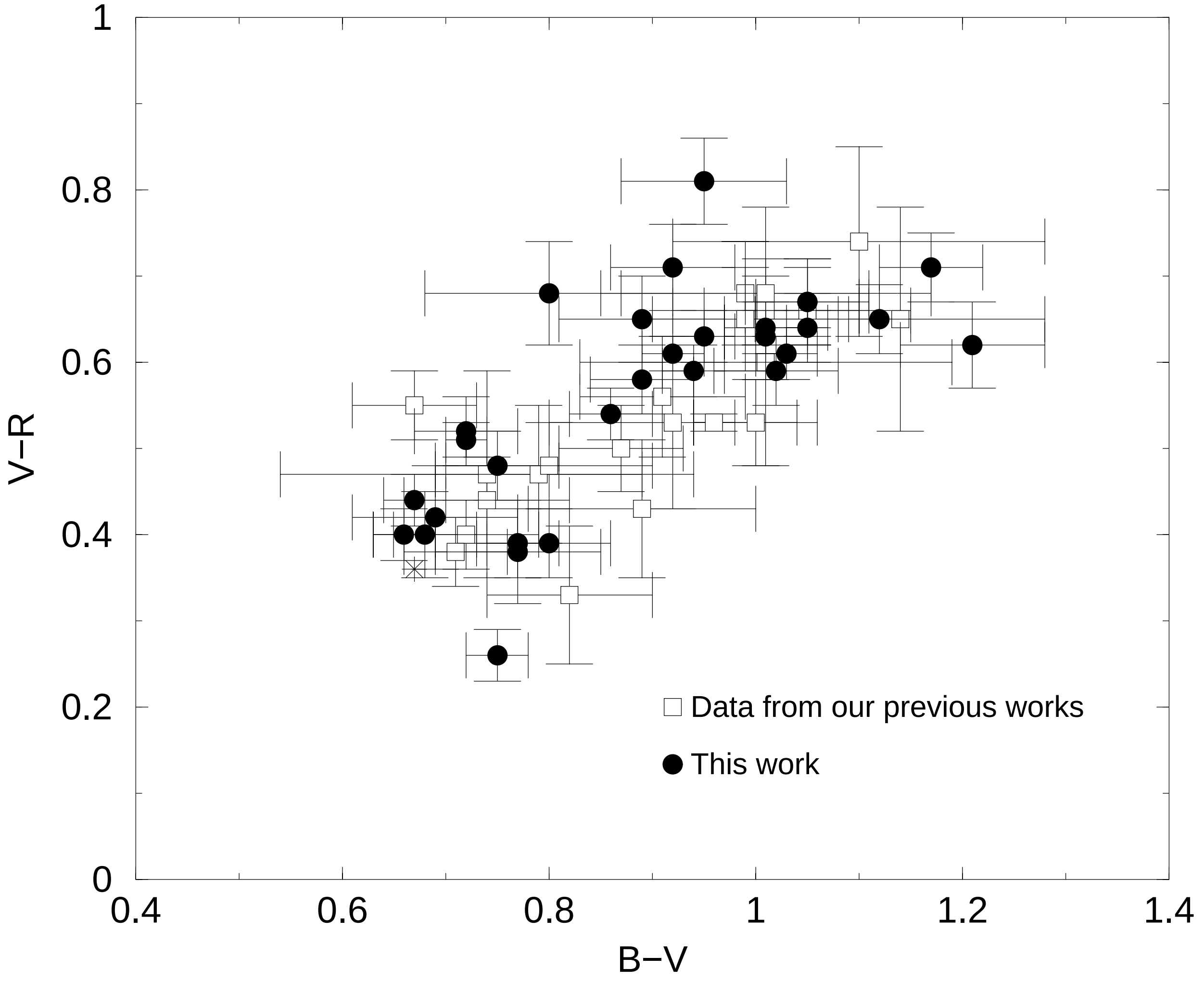

**Figure 2**

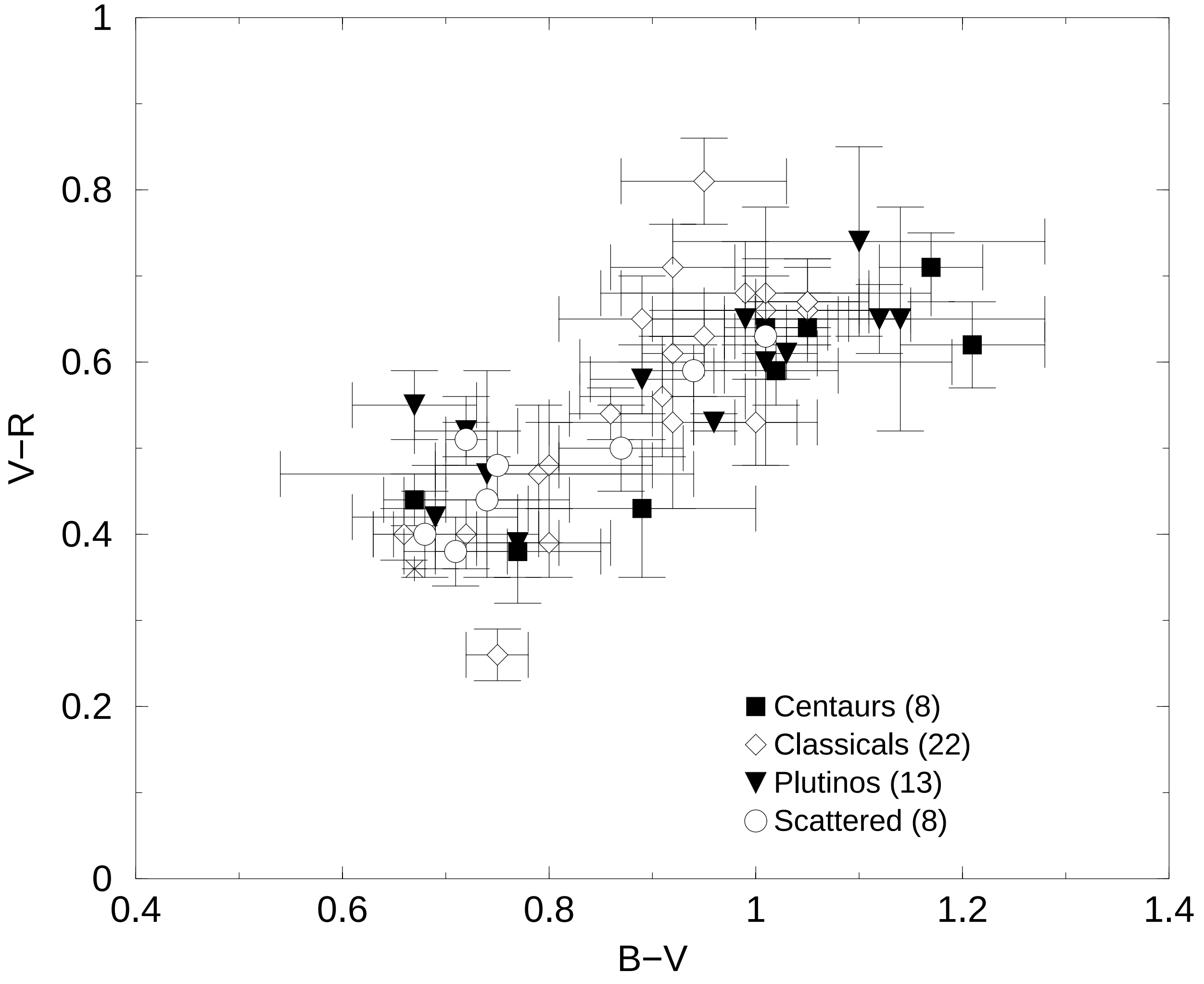

**Figure 3**

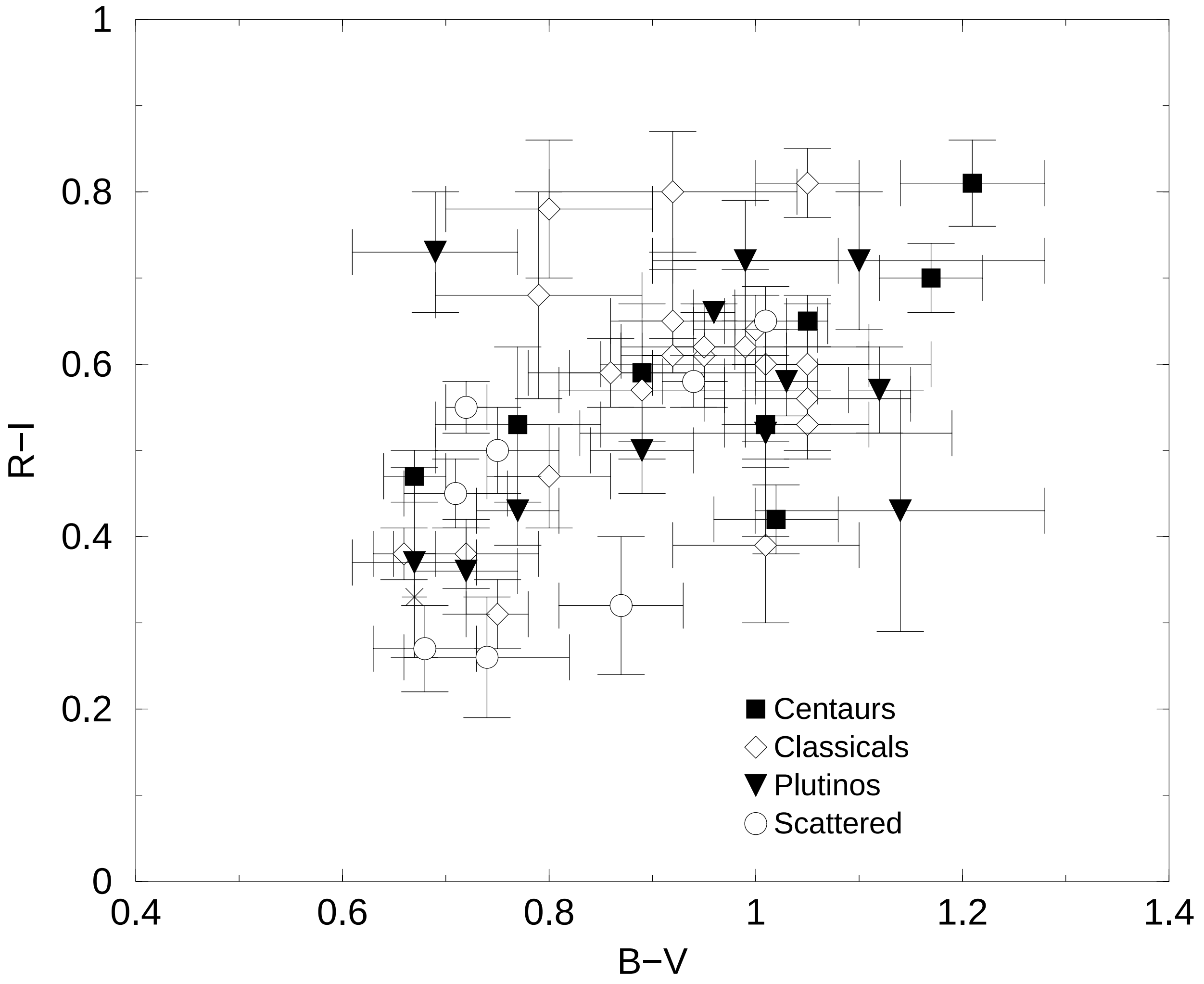

**Figure 4**

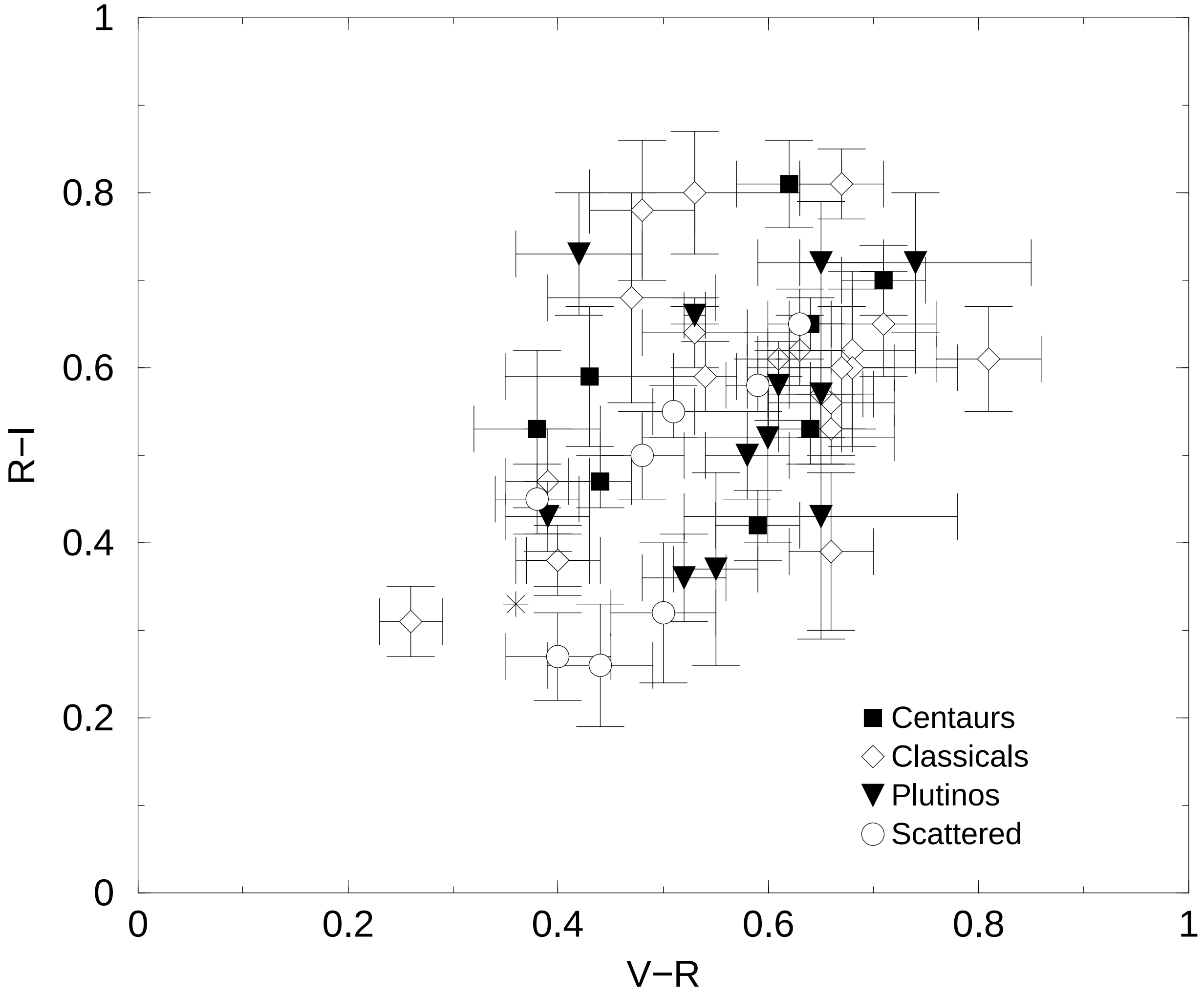

**Figure 5**

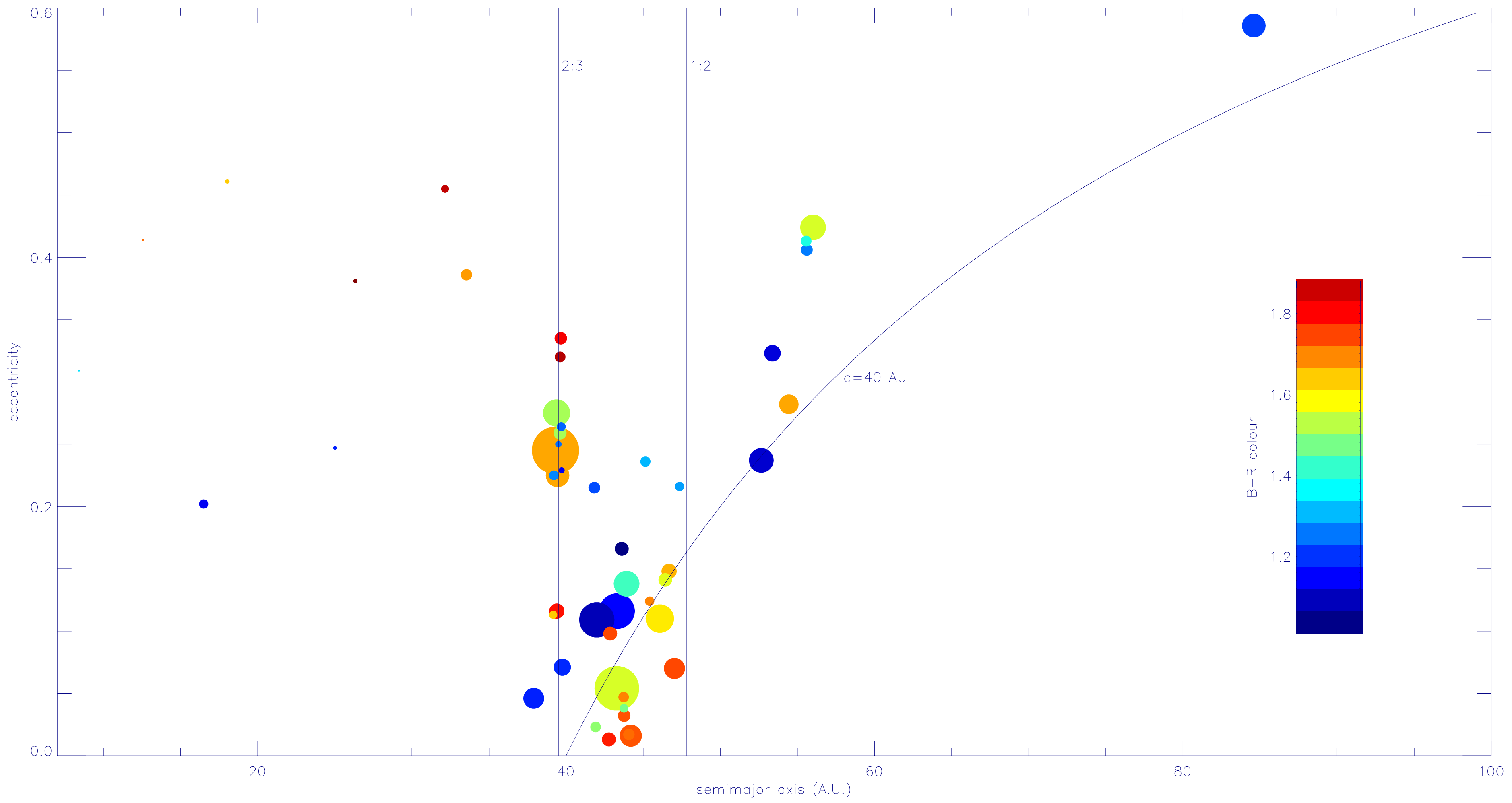

**Figure 6**

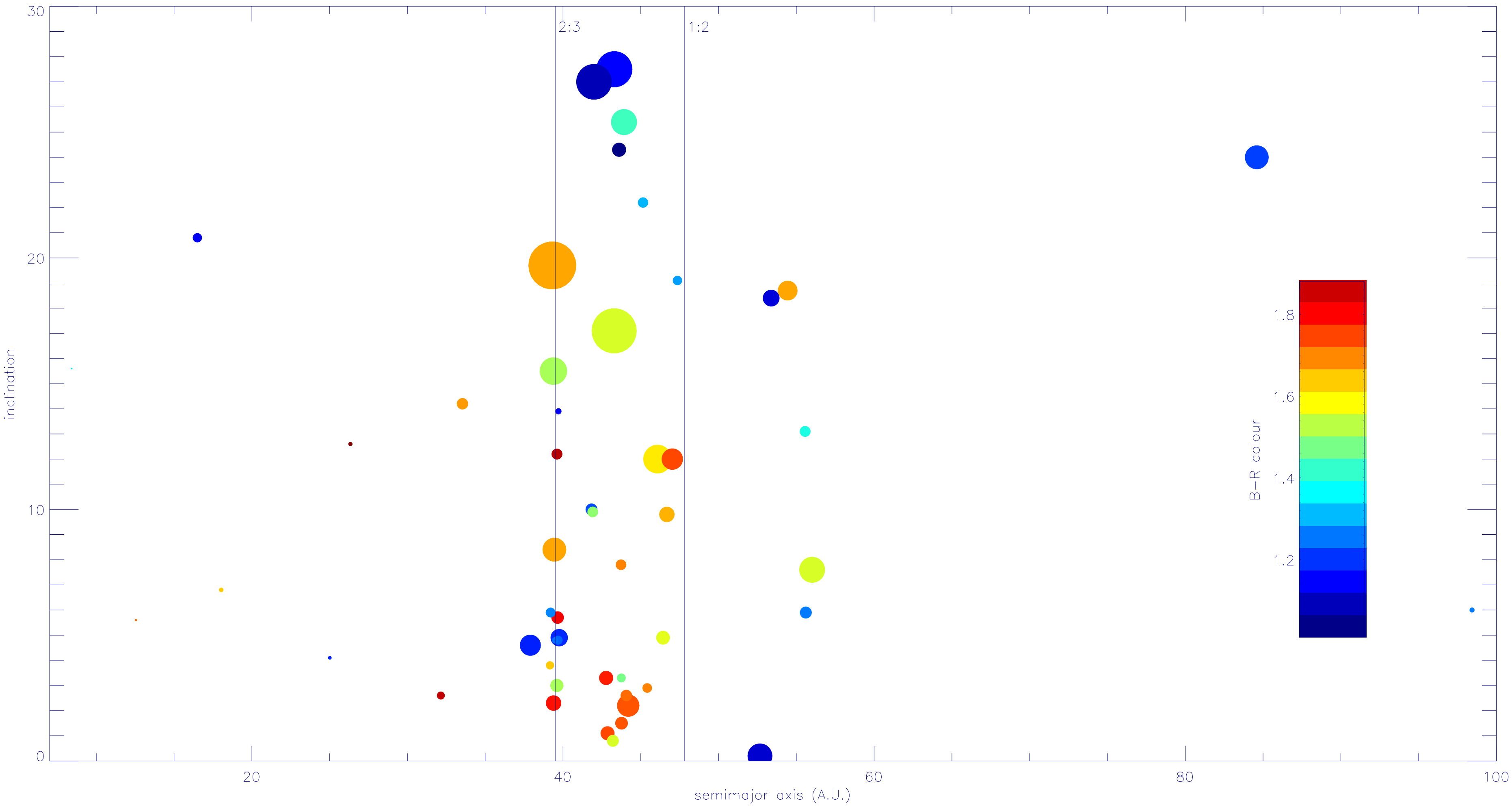

**Figure 7**

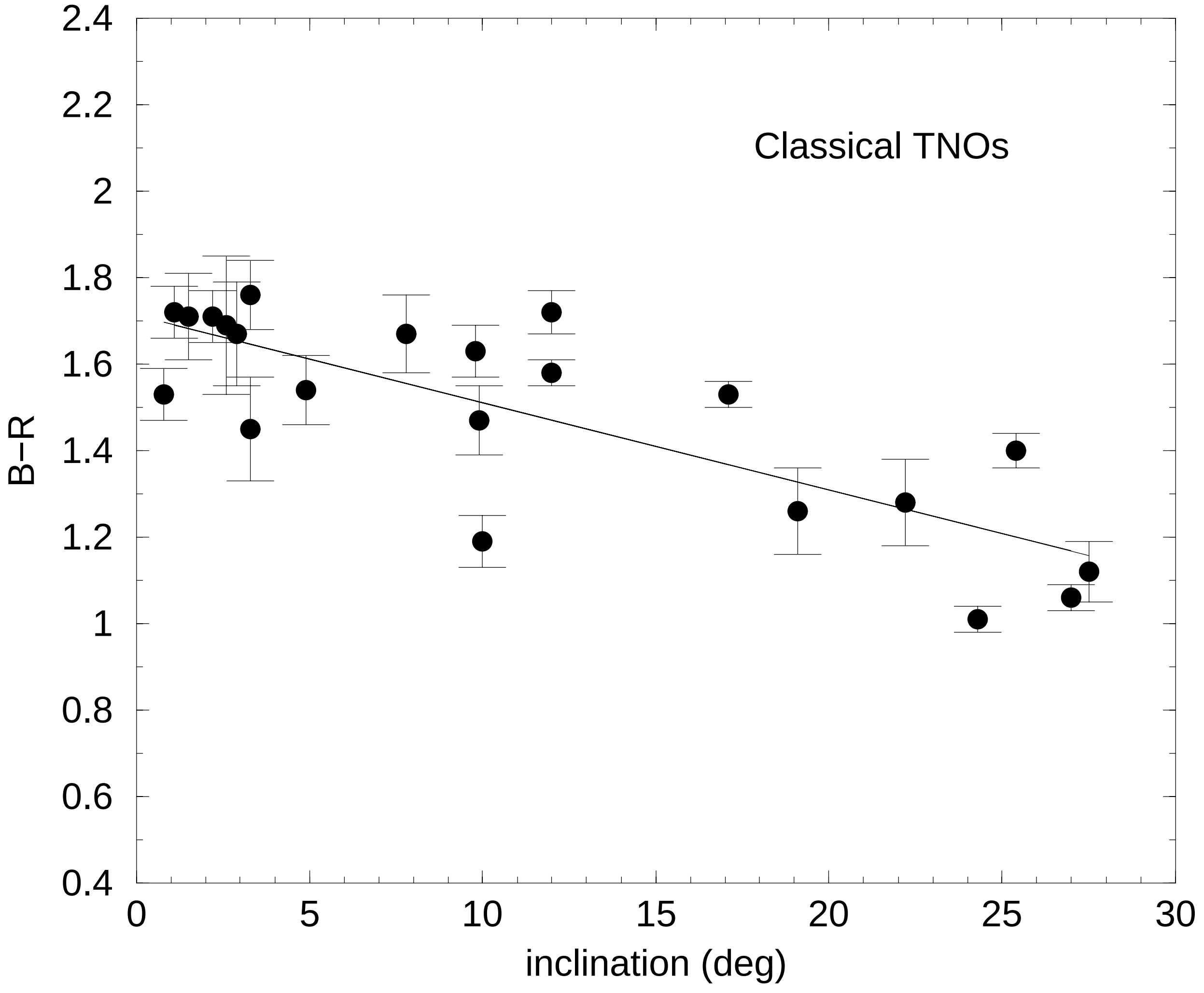

**Figure 8**

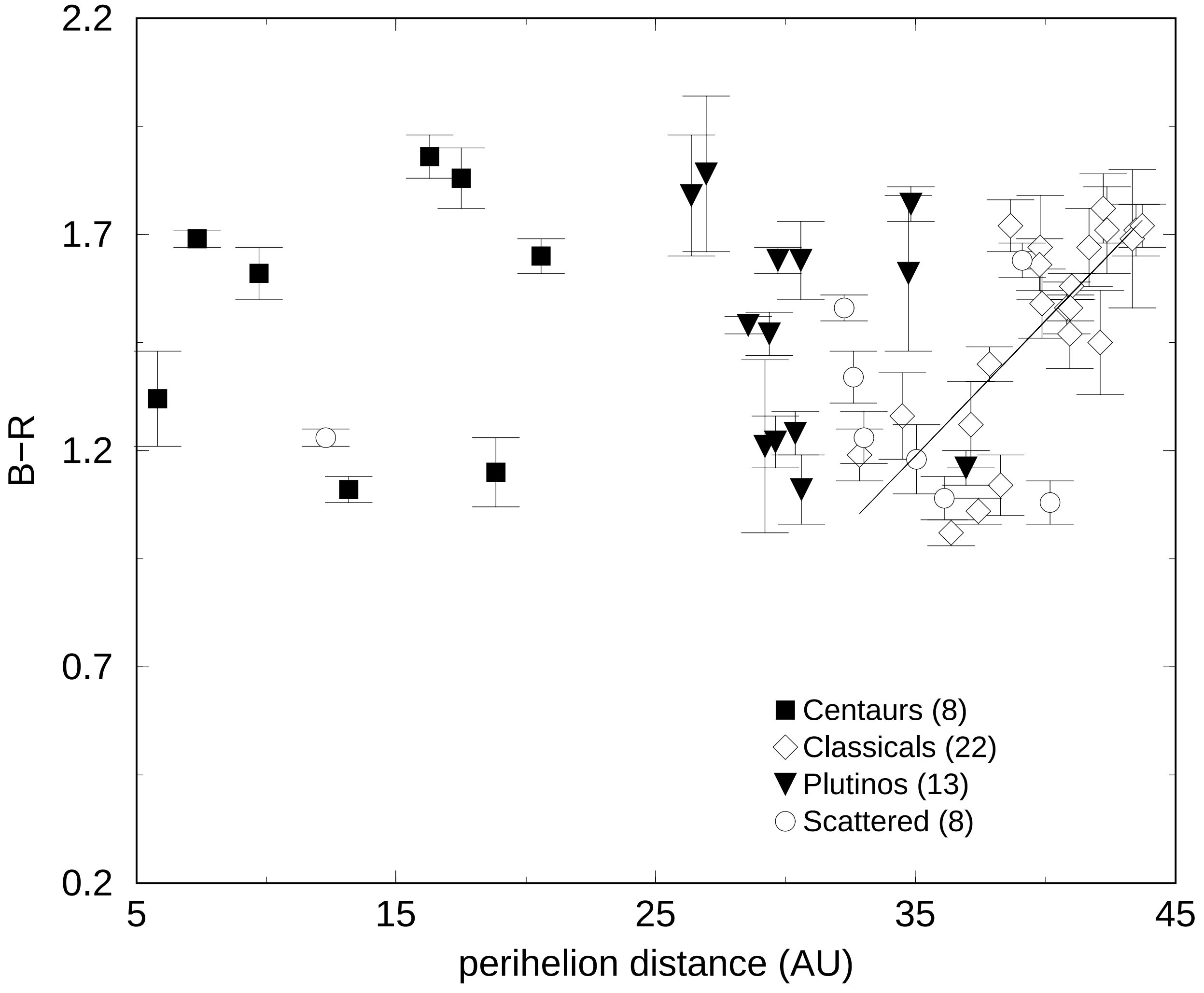

**Figure 9**

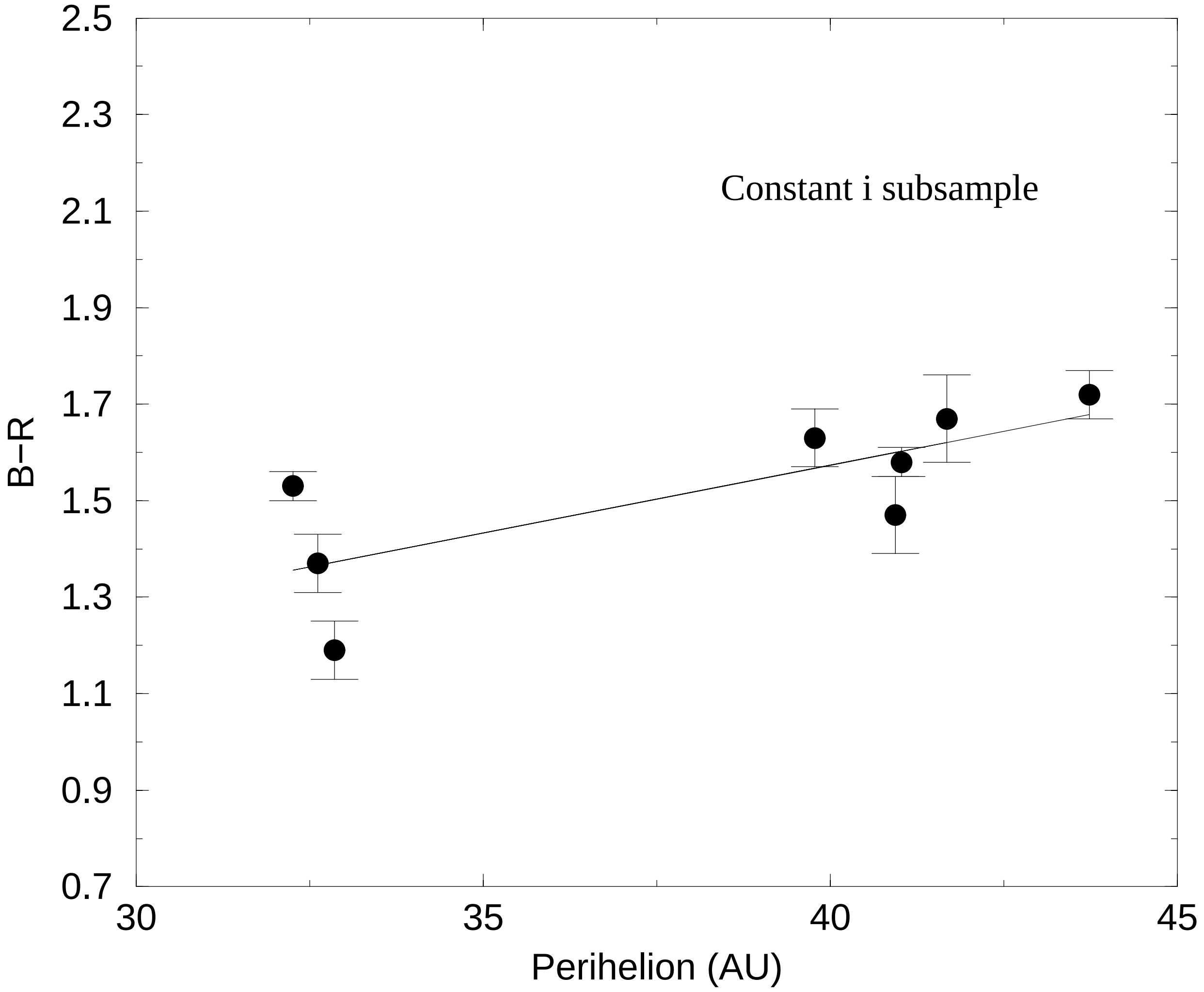

**Figure 10**

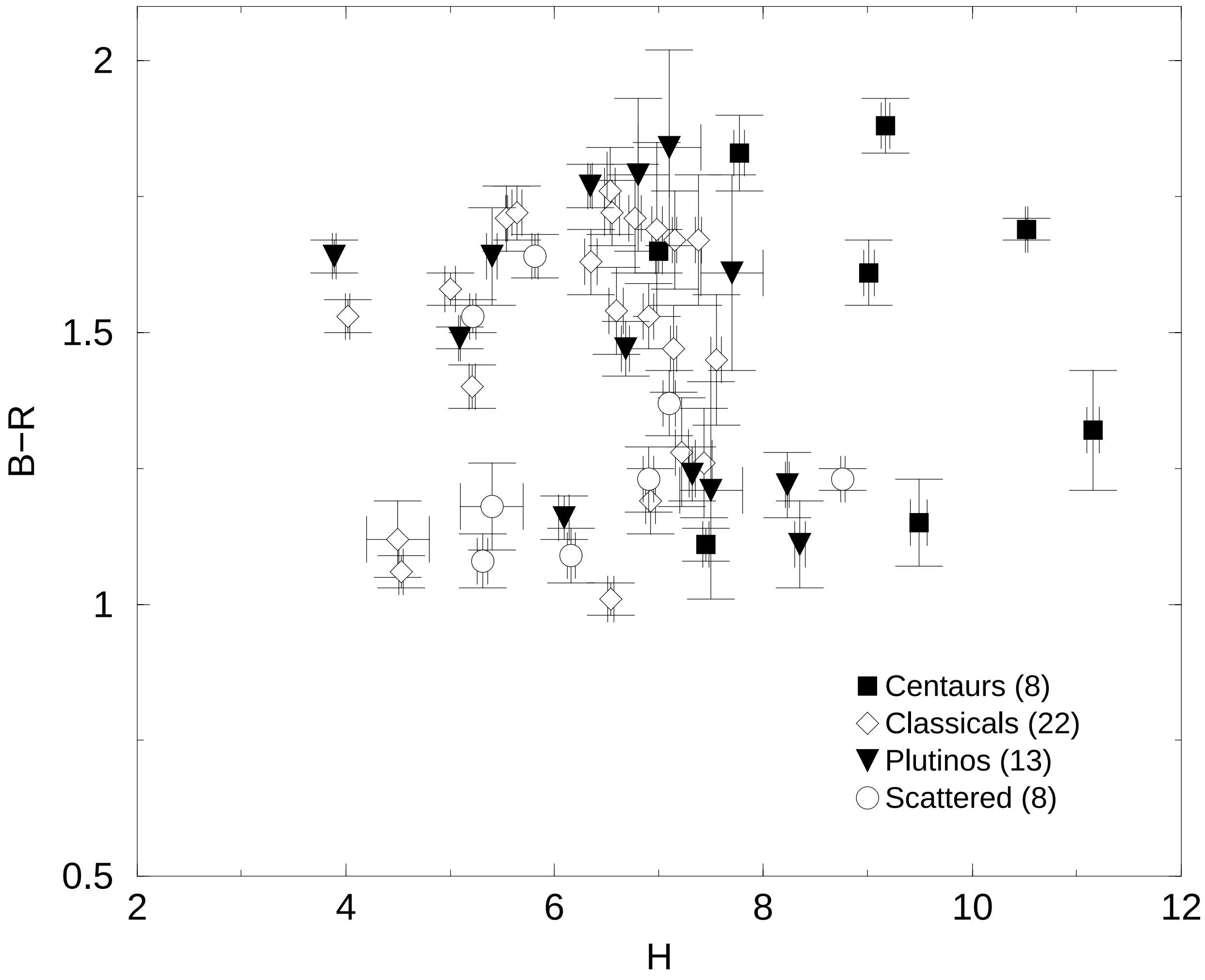

**Figure 11**

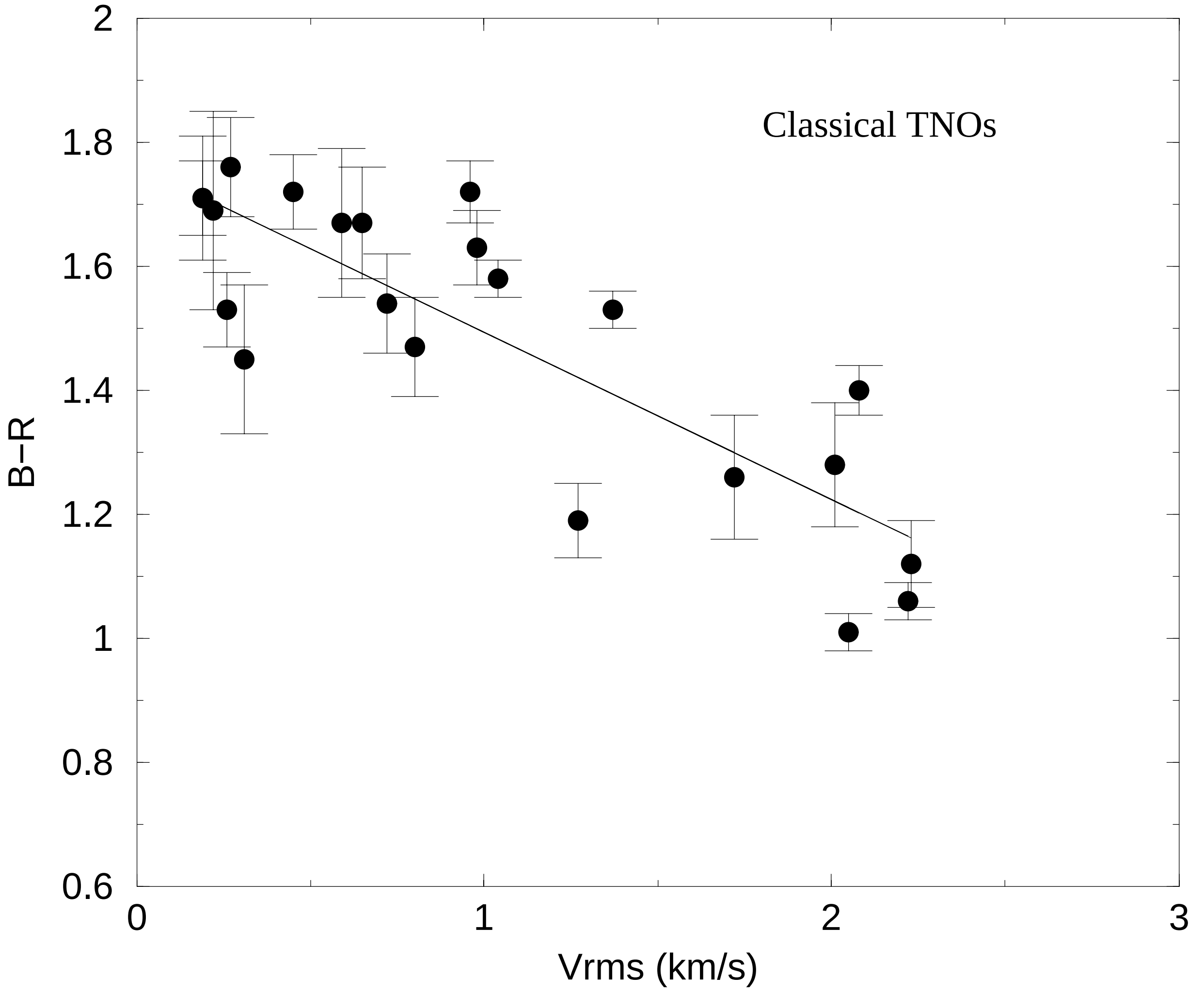